\documentclass[twocolumn, resetfootnote]{aastex701}

\usepackage{nameref}
\usepackage{amsmath}
\usepackage{graphicx}
\usepackage{float}
\usepackage{xcolor}
\usepackage{natbib}

\usepackage{multirow, array, makecell, booktabs, threeparttable} 

\usepackage{enumitem}
\setlist[itemize]{align=parleft}

\usepackage{tikz}
\usetikzlibrary{shapes.geometric, arrows, positioning}
\tikzstyle{startstop} = [rectangle, rounded corners, minimum width=3cm, minimum height=1cm, text centered, draw=black, fill=gray!20]
\tikzstyle{process} = [rectangle, minimum width=3cm, minimum height=1cm, text centered, draw=black]
\tikzstyle{decision} = [diamond, aspect=2, minimum width=3cm, minimum height=1cm, text centered, draw=black]
\tikzstyle{arrow} = [thick,->,>=stealth]

\received{August 8, 2025}
\revised{February 19, 2026}
\accepted{March 10, 2026}

\begin{document}

\title{Oxygen left behind: Atmospheric Enrichment due to Fractionation in Sub-Neptunes using \textsc{BOREAS}
\footnote{Released on \today}}

\author[0009-0005-3682-1593]{Marilina Valatsou}
\email{mvalatsou@phys.ethz.ch}
\affiliation{Institute for Particle Physics and Astrophysics, ETH Zurich, Wolfgang-Pauli-Strasse 27, 8093 Zurich Switzerland}

\author[0000-0001-6110-4610]{Caroline Dorn}
\email{dornc@phys.ethz.ch}
\affiliation{Institute for Particle Physics and Astrophysics, ETH Zurich, Wolfgang-Pauli-Strasse 27, 8093 Zurich Switzerland}

\author{Pierlou Marty}
\email{test}
\affiliation{Institute for Particle Physics and Astrophysics, ETH Zurich, Wolfgang-Pauli-Strasse 27, 8093 Zurich Switzerland}

\author[0000-0002-4856-7837]{James E. Owen}
\email{test}
\affiliation{Astrophysics Group, Imperial College London, Prince Consort Road, London SW7 2AZ, UK}


\begin{abstract}
The evolution of exoplanetary atmospheres is strongly influenced by atmospheric escape, particularly for close-in planets. Fractionation during atmospheric loss can preferentially remove lighter elements such as hydrogen, while retaining heavier species like oxygen. In this study, we investigate how and under what conditions hydrodynamic escape and chemical fractionation jointly shape the mass and composition of exoplanet atmospheres, especially for mixed H$_2$+H$_2$O atmospheres. We develop \texttt{BOREAS}, a self-consistent mass loss model coupling a one-dimensional Parker wind formulation with a mass-dependent fractionation scheme, which we apply across a range of planet masses, radii, equilibrium temperatures, and incident X-ray and ultraviolet (XUV) fluxes, allowing us to track hydrogen and oxygen escape rates at different snapshots in time. We find that oxygen is efficiently retained over most of the parameter space. Significant oxygen loss occurs under high incident XUV fluxes, while at intermediate fluxes oxygen loss is largely confined to low-gravity planets. Where oxygen is retained, irradiation is too weak to drive significant escape of hydrogen and thus limiting atmospheric enrichment. By contrast, our model predicts that sub-Neptunes undergo substantial atmospheric enrichment over $\sim$200 Myr when hydrogen escape is efficient and accompanied by partial oxygen entrainment. Notably, our results imply that sub-Neptunes near the radius valley can evolve into water-rich planets, in agreement with GJ 9827 d. Present-day water-rich atmospheres may have originated from water-poor envelopes under some conditions, highlighting the need to include chemical fractionation in evolution models. \textsc{BOREAS} is publicly available.
\end{abstract}




\section{Introduction} \label{sec:intro}
The rapid increase in exoplanet detections over the past two decades has dramatically improved our understanding of planetary formation, evolution, and atmospheric dynamics. Among the most striking features of the exoplanet population is the wide diversity of atmospheric compositions, spanning from hydrogen-dominated primordial envelopes to high-meanmolecular-weight ones, and are consistent with substantial volatile processing or secondary origins in some systems \citep[e.g.,][]{Moses2013-qx, Kempton2023-hm}. This diversity is widely interpreted---especially for close-in planets---as the outcome of multiple evolutionary pathways, including primordial gas accretion, secondary atmosphere production through volatile outgassing, and atmospheric mass-loss. The latter process is driven by intense stellar radiation, and can alter atmospheric composition, surface conditions, and longterm habitability \citep{owen_atmospheric_2019, tian_atmospheric_2015}. Volatile loss due to irradiation is also believed to have sculpted the terrestrial planets of our own solar system, as demonstrated by isotopic and noble-gas signatures for Venus, Earth, and Mars \citep{lammer_atmospheric_2008}.

Observations of sub-Neptunes suggest a range of possibilities: Some appear as steam worlds \citep{piaulet-ghorayeb_jwstniriss_2024}, while others may retain hydrogen-dominated envelopes, such as TOI-270 d \citep{eylen_masses_2021} and K2-18 b \citep{madhusudhan_carbon-bearing_2023}. Although there is considerable debate over whether sub-Neptunes are predominantly gas dwarfs or water worlds \citep{bean_nature_2021}, recent global chemical equilibrium models indicate that all sub-Neptunes likely form water-poor, with bulk water mass fractions (WMFs) $<2$\%, and initial water envelope mass fractions below 20\% \citep{werlen_sub-neptunes_2025}. Taken together, these results suggest that fractionated mass-loss during a planet’s evolution may drive the atmospheric diversity of subNeptunes, since lighter species like hydrogen escape more readily than heavier ones like oxygen. Here, we explore this hypothesis. Loss of volatiles such as water and hydrogen is crucial in determining whether a planet keeps its atmosphere or becomes a bare rocky world \citep{luger_extreme_2015}.

Hydrodynamic escape is a primary driver of atmospheric mass-loss, particularly in young planetary systems, where high-energy stellar radiation ionizes atmospheric gases and powers outflows that can strip hydrogen-rich envelopes and transform sub-Neptunes into super-Earths over geological timescales \citep{lopez_how_2012, jin_planetary_2014, rogers_photoevaporation_2021}. The efficiency of this escape depends on stellar irradiation, planetary gravity, and atmospheric thermodynamics, with regimes such as energy- and recombination-limited escape setting mass-loss rates \citep{murray_atmospheric_2009, owen_uv_2015}.

Both photoevaporation and core-powered mass-loss shape the radius distribution of super-Earths and sub-Neptunes. Rather than competing, they act sequentially depending on system evolution and energy budget: Photoevaporation is driven by stellar X-ray and ultraviolet (XUV) radiation \citep{lammer_atmospheric_2003}, while core-powered loss relies on residual cooling luminosity \citep{ginzburg_core-powered_2018}. Which process dominates depends on the sonic point’s location relative to the XUV absorption depth. Population models reproduce the observed radius valley, highlighting photoevaporation as the primary sculptor, while core-powered loss remains important for weakly bound, highly irradiated planets \citep{owen_mapping_2023}, though \citet{loyd_current_2020} find neither mechanism universally dominant. Here, we adopt a photoevaporation-dominated framework and do not explicitly model core-powered mass-loss, although we can identify such outflows within out modeling framework.

In addition to mass-loss, fractionation plays a key role in shaping atmospheric composition over time. For example, high-energy stellar radiation can dissociate the water molecules of a steam atmosphere into hydrogen and oxygen atoms, whose escape is governed by diffusive separation and hydrodynamic drag
\citep{zahnle_mass_1986}. Because hydrogen escapes more easily than oxygen, the atmosphere may become enriched in heavier elements---a trend supported by studies of transonic outflows in the Solar System \citep{hunten_mass_1987, zahnle_mass_1990} and recent exoplanet models. \citet{cherubim_strong_2024} showed that the degree of fractionation is highly dependent on the escape mechanism, with photoevaporation and core-powered loss producing distinct signatures in deuterium and helium. Preferential hydrogen loss can even cause abiotic O2 buildup, a possible false biosignature as shown by \citet{wordsworth_water_2013, luger_extreme_2015, schaefer_predictions_2016, wordsworth_redox_2018}, though other models suggest oxygen is largely dragged out with hydrogen \citep{burn_water-rich_2024}. The application of fractionated loss at the population level \citet{cherubim_oxidation_2025} has already revealed strong oxidation gradients and species-dependent loss signatures, demonstrating that coupled fractionation-escape modeling is feasible at scale. Yet most models treat mass-loss and fractionation separately \citep{cherubim_oxidation_2025}, leaving key questions unresolved. Full radiation hydrodynamic models could address this but remain computationally challenging for large populations. 

Our approach differs from previous developments as follows: (1) We compute the wind structure using a semi-analytic hydrodynamic solution by \citet{owen_mapping_2023}, further built upon by \citet{ballabio_understanding_2025}, that self-consistently solves for the sonic point, momentum balance, and XUV absorption radius, and we determine the transition between energy-limited and recombination-limited escape from the wind solution itself, rather than adopting analytic limiting-flux scalings. (2) We couple this wind solution to a multicomponent fractionation scheme in which the mean molecular weight, sound speed, XUV radius, and species fluxes (H and O, or more generally light and heavy atoms) are iterated to convergence. This contrasts with IsoFATE \citep{cherubim_strong_2024}, where fractionation is applied as a postprocessing step to a fixed hydrodynamic escape rate. (3) We target a distinct atmospheric regime: pure steam, pure H/He, and mixed H$_2$+H$_2$O envelopes (10–90\% water by mass). This allows explicit O/H fractionation in water-rich atmospheres, while IsoFATE focuses on D/H and He/H fractionation in primarily H/He envelopes with water playing only a catalytic role.

Our semi-analytic hydrodynamic framework combines a generalized isothermal Parker-wind approach with self-consistent XUV penetration and recombination-limited physics to enable efficient exploration of a broad parameter space. Unlike simple energy-limited scaling models, where the XUV absorption radius is often assumed to match the planet’s radius, our method calculates the XUV radius; these are also the calculations that allow us to determine the transition from core-powered mass-loss to photoevaporation. Additionally, it incorporates ionization-recombination equilibrium where appropriate, and enforces a matching condition for the sound speed and mass-loss rate without requiring full radiation-hydrodynamic simulations. At the same time, it simplifies certain aspects, such as the photodissociation efficiency and the detailed energy deposition, compared to full radiation-hydrodynamic codes.

We extend our hydrodynamic formulation with a fractionation scheme grounded in the classical diffusion–drag framework of \citet{hunten_mass_1987} and \citet{zahnle_mass_1986}, later refined by \citet{zahnle_mass_1990} and \citet{zahnle_photochemistry_1986}. In our implementation, the coupled escape of light and heavy species is solved self-consistently with the total mass-loss rate, rather than assuming a purely diffusion-limited flux, where hydrogen escape is throttled by its ability to diffuse through heavier molecules. Instead, the species fluxes respond to the strength of the hydrodynamic wind, allowing us to capture both the regime in which heavy atoms are efficiently dragged along by the bulk outflow and the regime in which their escape becomes diffusion limited. This approach follows the spirit of the multicomponent hydrodynamic models of \citet{odert_escape_2018}, but remains semi-analytic and computationally efficient.

We explore simulations for a population of planets representative of the observed super-Earth and sub-Neptune population. Lastly, we benchmark our coupled model against existing mass-loss rate estimates for well-characterized exoplanet systems (see Appendix~\ref{sec:comparison}). The detailed description of the photoevaporation framework is provided in Section~\ref{sec:escape_model}, and the fractionation methodology is outlined in Section~\ref{sec:fractionation_model}, with the results of applying it to super-Earth and sub-Neptune environments in Section~\ref{sec:results}.
\section{Atmospheric Escape Model} \label{sec:escape_model}
We build an XUV-driven photoevaporation model based on the modified Parker wind framework of \citet{owen_mapping_2023}. We implement this framework in our new, open-source code \textsc{BOREAS}\footnote{\url{https://github.com/ExoInteriors/BOREAS}},which self-consistently solves the hydrodynamic structure and species-dependent escape for the parameter space explored in this work. This approach accounts for both energy-limited and recombinationlimited escape regimes, which depend on the atmospheric density and irradiation conditions. In the energy-limited regime, escape is driven by XUV heating, while in the recombination-limited regime, intense ionization forces hydrogen in a dense flow and radiative recombination becomes the main cooling mechanism. This recombination cooling thermostat sets the outflow to a temperature of approximately 10$^4$ K and thus caps the mass-loss rate. The regime classification and diagnostics follow \citet{owen_mapping_2023}. The mass-loss framework is coupled with a stellar XUV model that builds on \citet{baraffe_new_2015} and \citet{rogers_photoevaporation_2021}. 

\subsection{Stellar Irradiation Model} \label{sec:stellar_flux}
\begin{figure}[h]
    \centering
    \includegraphics[width=\linewidth]{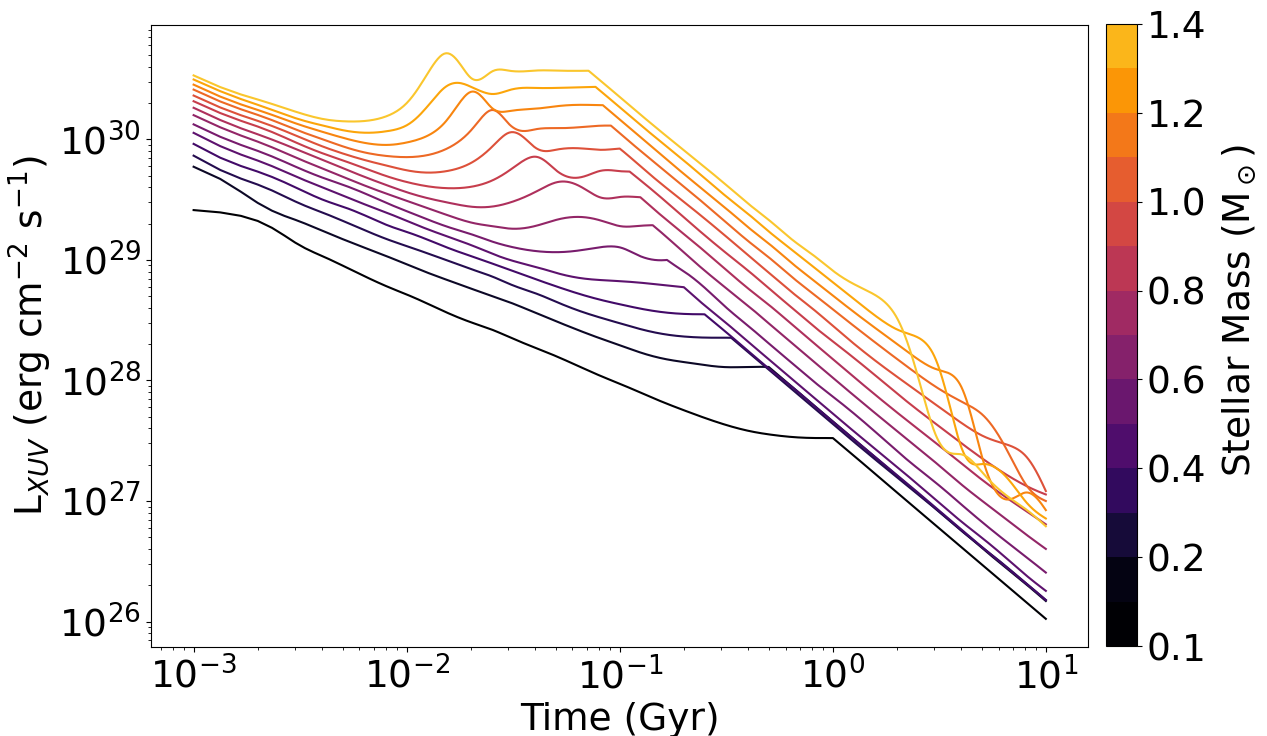}
    \caption{Time evolution of stellar XUV luminosity for stars of different masses, ranging from 0.1 to 1.4 M$_\odot$.}
    \label{fig:LXUV_evolution}
\end{figure}

To calculate the incident XUV flux $F_\mathrm{XUV}$ for a planet, we first estimate the star’s XUV luminosity $L_\mathrm{XUV}$. We assume that stars follow a characteristic ratio $L_\mathrm{XUV}/L_\mathrm{bol}$, which we interpolate as a function of stellar mass and age using the evolutionary tracks of \citet{baraffe_new_2015} and the empirical relations compiled in \citet{rogers_photoevaporation_2021}. Although this ratio varies weakly with stellar mass at fixed age, it captures the essential trend that younger and more massive stars emit higher XUV radiation than older or lower-mass stars. The resulting time evolution of $L_\mathrm{XUV}$ for different stellar masses is shown in Figure~\ref{fig:LXUV_evolution}. Once $L_\mathrm{XUV}$ is determined, the incident flux $F_\mathrm{XUV}$ at the planet’s orbit is calculated directly from this luminosity and the orbital distance.

To establish the relevant orbital distance, we link the planet’s equilibrium temperature $T_\mathrm{eq}$ to the star’s bolometric luminosity $L_\mathrm{bol}$ using the standard planetary energy balance relation, which equates the absorbed stellar flux to the planet’s thermal re-radiation. Specifically, we assume a Bond albedo $\alpha = 0.3$, a re-radiation fraction $\beta = 0.75$, and an emissivity $\epsilon = 1$. Here, $\beta$ specifies what fraction of the absorbed stellar flux is re-radiated across the planet’s surface. A value of $\beta = 1$ corresponds to full heat redistribution across the entire sphere, while $\beta = 0.5$ corresponds to re-radiation from only the dayside. We adopt a value for an intermediate case suitable for planets with partial day–night heat redistribution. The emmisivity $\epsilon$ denotes the efficiency with which the atmosphere radiates thermal energy to space. We assume unit emissivity corresponding to a blackbody emitter, consistent with standard equilibrium-temperature calculations. While these parameters are fixed for the models presented here, they remain configurable inputs to our framework for future analyses.

Finally, we adopt a heating efficiency of $\eta = 0.3$, which represents the fraction of absorbed XUV energy that effectively drives atmospheric escape rather than being radiated away from the planet. This value is consistent with prior photoevaporation models \citep{owen_mapping_2023} and is found to be appropriate for the low density planets we consider in this work \citep{owen_planetary_2012, kubyshkina_grid_2018}.

\subsection{Hydrodynamic Escape Approach} \label{sec:hydro_eqs}
\begin{figure}
    \centering
    \includegraphics[width=\linewidth]{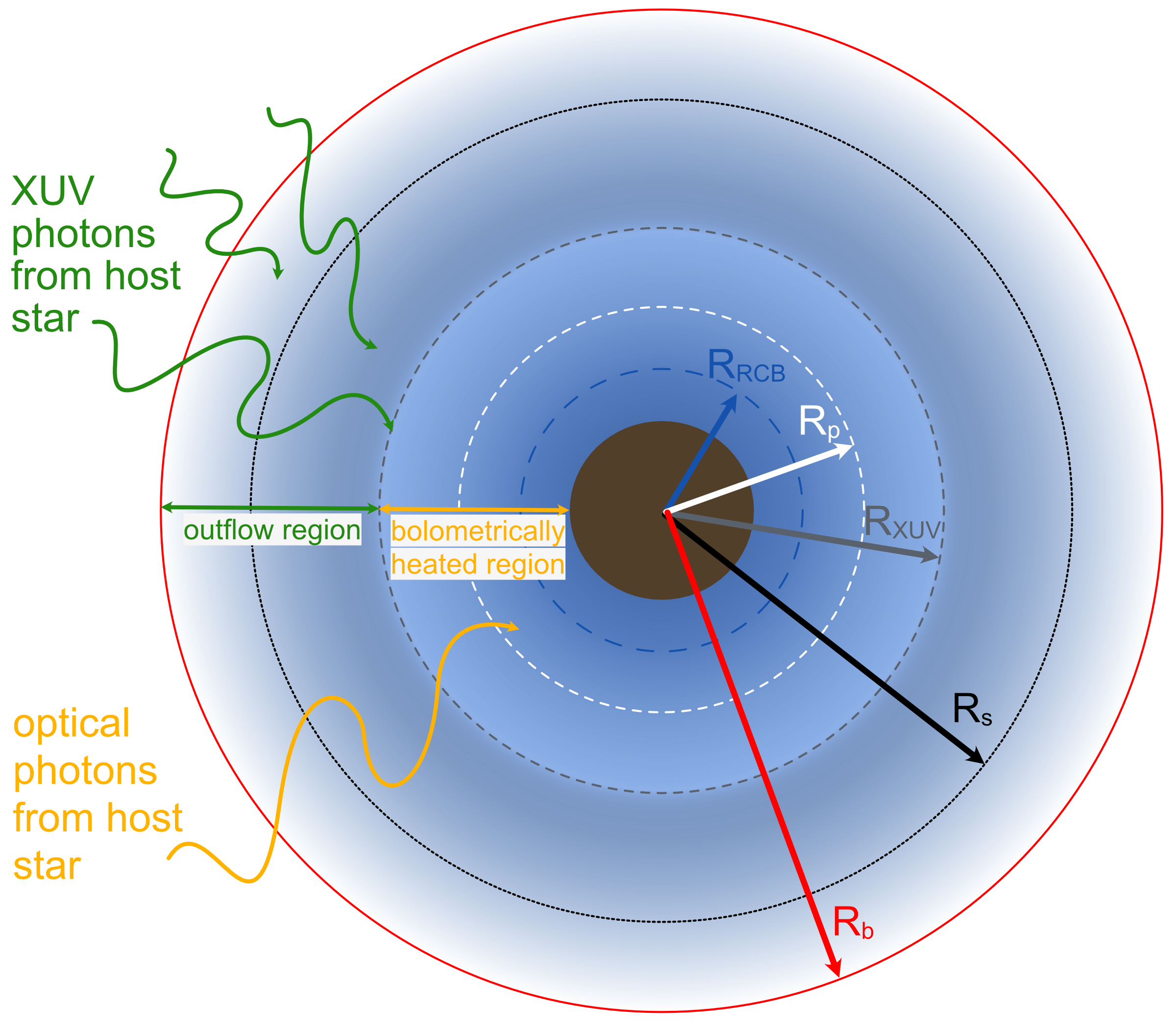}
    \caption{Schematic depiction of the model's atmosphere, adapted from \cite{owen_mapping_2023}. For a planet undergoing significant photoevaporation, $R_B$ lies outside $R_\mathrm{XUV}$, and $R_{s} < R_B$.}
    \label{fig:model}
\end{figure}

We model atmospheric outflow using a 1-dimensional, spherically symmetric, steady-state Parker wind model following the methodology of \citet{owen_mapping_2023}. The general structure of the modeled atmosphere used to calculate mass-loss is illustrated in Figure~\ref{fig:model}.

We approximate the radiative region above the radiative-convective boundary $R_\mathrm{rcb}$ as isothermal at the planet’s $T_\mathrm{eq}$, extending outward to the XUV absorption radius $R_\mathrm{XUV}$. This treatment follows prior work on weakly luminous irradiated atmospheres \citep{owen_atmospheres_2016, owen_evaporation_2017} and is consistent with the tendency of radiative-equilibrium temperature profiles to flatten once stellar irradiation has been absorbed in semi-grey atmospheric models \citep{guillot_radiative_2010}. The interior-structure model used in this study also follows the same method (see Section~\ref{sec:models_params}). Beyond $R_\mathrm{XUV}$, stellar XUV photons heat the gas to higher temperatures $T_\mathrm{outflow}$, driving a photoevaporative outflow. The outflow remains subsonic from the base of the wind up to the sonic point at $R_{s}$, beyond which it transitions to supersonic expansion.

Within this structure, at smaller radii than $R_\mathrm{XUV}$, we also define the transit radius of the planet $R_\mathrm{transit}$. This is the radius at which the atmosphere becomes optically thick in transit photometry and it corresponds to the photospheric radius $R_p$ (e.g., in Eq.~\ref{app:rho_XUV}). The photospheric radius is equal to the modeled planetary radius $R_{p}$ (see Section~\ref{sec:models_params}), thus $R_p = R_\mathrm{transit}$, which is a good approximation for evolved planets but not necessarily young puffy worlds. Finally, the Bondi radius $R_{B}$ associated with a locally isothermal outflow is:

\begin{equation}
   R_B = \frac{G M_p}{2 c_{s}^2},
   \label{eq:bondi_radius}
\end{equation}

where $c_s$ is the local sound speed, and $M_{p}$ is the planetary mass. For significant photoevaporation to occur in the sense of \citet{owen_mapping_2023}, we generally require $R_\mathrm{XUV} < R_{s} < R_B$, such that XUV heating occurs inside the subsonic region and the wind can pass smoothly through the sonic point. In scenarios where $R_\mathrm{XUV} > R_{s}$, mass-loss can still occur; we treat these cases using the modified Parker-wind branch discussed in Appendix~\ref{app:hydro_eqs}. However, if $R_\mathrm{XUV} > R_B$, this scenario would correspond to a core-powered mass-loss regime. As none of our explored parameter space falls within this regime, we do not model it here.

\subsection{Numerical Scheme and Density Structure} \label{sec:numerics}
We discretize the radial domain (typically on a logarithmic grid) from $R_{p}$ outward to several planetary radii, e.g., $5 R_p$, covering the region affected by XUV-driven escape. The hydrodynamic solution is obtained by coupling an isothermal Parker wind above $R_\mathrm{XUV}$ to a hydrostatic, isothermal layer below $R_\mathrm{XUV}$. For a given trial $R_\mathrm{XUV}$, we determine an outflow sound speed $c_s$ such that the resulting wind power is consistent with energy-limited heating. Specifically, the energy-limited mass-loss rate is

\begin{equation}
    \dot{M}_\mathrm{EL} = \eta \, F_\mathrm{XUV}\,\frac{\pi R_\mathrm{XUV}^3}{4\,G\,M_p},
\end{equation}


where $F_\mathrm{XUV}$ is the incident XUV flux, and the factor of 1/4 accounts for the planet's cross-sections relative to its surface area when computing the globally averaged flux. For fixed $R_\mathrm{XUV}$, we solve for $c_s$ using a root-finding procedure (Brent's method) such that the Parker-wind solution yields $\dot{M}=\dot{M}_\mathrm{EL}$. The instantaneous mass-loss rate is computed from the wind base as

\begin{equation}
    \dot{M} = 4\pi R_\mathrm{XUV}^2 \,\rho(R_\mathrm{XUV})\,u(R_\mathrm{XUV}),
\end{equation}

where $u(R_\mathrm{XUV})$ and $\rho(R_\mathrm{XUV})$ are the velocity and density at the XUV base.

In our implementation, $\rho(R_\mathrm{XUV})$ is set by requiring that the XUV optical depth above $R_\mathrm{XUV}$ be of order unity using a mass absorption coefficient $\chi_\mathrm{XUV}$ (units $\mathrm{cm^2\,g^{-1}}$). For a given Parker-wind velocity field $u(r)$, we write the density profile up to an overall scale as $\rho(r)\propto (R_s/r)^2\,(c_s/u(r))$ and compute the 
dimensionless column integral of the wind density profile

\begin{equation}
    \tau_\mathrm{geom}(R_\mathrm{XUV}) \equiv \int_{R_\mathrm{XUV}}^\infty \left(\frac{R_s}{r}\right)^2 \left(\frac{c_s}{u(r)}\right)\,dr,
\end{equation}

so that the density normalization follows from the condition $\tau_\mathrm{XUV}\sim 1$, yielding $\rho(R_\mathrm{XUV}) = (\chi_\mathrm{XUV}\,\tau_\mathrm{geom})^{-1}$.

Finally, the physically consistent $R_\mathrm{XUV}$ is obtained by enforcing momentum balance across the transition from the hydrostatic layer to the photoevaporative flow, following \citet{owen_mapping_2023}. Concretely, for each trial $R_\mathrm{XUV}$ we compare (i) the hydrostatic pressure support implied by the bolometrically heated layer and (ii) the total momentum flux of the wind at launch. We scan over $R_\mathrm{XUV}$ and use root-finding to identify the $R_\mathrm{XUV}$ that satisfies the momentum-balance condition (Appendix~\ref{app:hydro_eqs}).

\subsection{Escape Rate Evaluation and Regime Classification} \label{sec:escape_rate}
Once an energy-limited solution is obtained, we evaluate whether the flow is instead recombination-limited using the diagnostic time-scale ratio \citep{owen_mapping_2023},

\begin{equation}
    \mathcal{R}_t \equiv \frac{t_\mathrm{rec}}{t_\mathrm{flow}} \;=\; \frac{t_\mathrm{rec}}{R_\mathrm{XUV}/u(R_\mathrm{XUV})},
\end{equation}

where $t_\mathrm{flow}\sim H/u(R_\mathrm{XUV})$ is the local advection time at the XUV base, $t_\mathrm{rec}\sim 1/(\alpha_\mathrm{rec} n_e)$ is the hydrogen recombination time, $\alpha_\mathrm{rec}$ is the case-B (recombinations happen to excited states) recombination coefficient and $n_e$ the electron density (see also \citealt{owen_mapping_2023}). When $\mathcal{R}_t>1$, recombination is slow compared to advection and the outflow remains in the energy-limited regime; in these cases we adopt $\dot{M}=\dot{M}_\mathrm{EL}$ for subsequent calculations. When $\mathcal{R}_t<1$, recombination is fast and the flow enters the recombination-limited regime. In that case we fix the outflow sound speed to $c_s \simeq 1.2\times10^6\,\mathrm{cm\,s^{-1}}$ (corresponding to $T\sim10^4$ K), re-solve for $R_\mathrm{XUV}$ using the recombination-limited closure, and adopt the resulting $\dot{M}$, $R_\mathrm{XUV}$, and $c_s$ as our recombination-limited solution, as indicated in the flowchart in Appendix~\ref{app:figures}, Figure~\ref{fig:flowchart}.

\section{Atmospheric Fractionation Approach} \label{sec:fractionation_model}
We couple the hydrodynamic escape model to a diffusion--drag fractionation framework originally developed for elemental and isotopic escape in terrestrial Solar System atmospheres (Venus, Earth, Mars) \citep{zahnle_mass_1986, hunten_mass_1987}, and implement it in the multi-species form described by \citet{odert_escape_2018}. In this work, we apply this approach specifically to hydrogen--oxygen fractionation in dissociated steam atmospheres and mixed H$_2$ + H$_2$O atmospheres. For any mixture, we assume complete dissociation of the major molecular reservoirs at and above $R_\mathrm{XUV}$ and therefore treat the escaping gas as an atomic mixture. In the present implementation, we neglect helium as an explicit third species in the fractionation network. Our two-component treatment does not resolve the homopause explicitly and assumes that eddy diffusion is sufficiently strong to maintain a well-mixed composition up to the base of the wind.

Because heavier species are less readily accelerated by hydrodynamic flows, the escaping gas composition may deviate from the atmospheric bulk composition. In our case, hydrogen (the light species) escapes more efficiently, whereas oxygen (the heavy species) may remain partially gravitationally bound, especially on planets with deep gravitational potential \citep{zahnle_mass_1986, hunten_mass_1987}. During episodes of significant atmospheric loss, this differential escape can alter the atmospheric composition and increase the O/H ratio in the remaining atmosphere. This directly affects the mean molecular weight of the escaping gas $\mu_\mathrm{outflow}$, which in turn influences key hydrodynamic escape quantities (e.g., density structure and the location of $R_\mathrm{XUV}$ through the XUV optical depth condition). To capture this feedback, our model embeds the fractionation calculation within the main iteration loop for mass-loss, ensuring that composition and outflow dynamics are solved jointly rather than in post-processing. The convergence procedure is detailed in Section~\ref{sec:self_consistent_fractionation}.

\subsection{Fractionation Factor and Flux Partitioning} \label{sec:frac}
Within each hydrodynamic iteration step, the mass-loss solver returns $(R_\mathrm{XUV}, c_s, \dot{M})$ and thus a total mass flux at the wind base,

\begin{equation}
    F_\mathrm{mass} \equiv \frac{\dot{M}}{4\pi R_\mathrm{XUV}^2}\qquad [\mathrm{g\,cm^{-2}\,s^{-1}}].
\end{equation}

We partition $F_\mathrm{mass}$ into \emph{number} fluxes $\phi_\mathrm{H}$ and $\phi_\mathrm{O}$ (units $\mathrm{cm^{-2}\,s^{-1}}$) for atomic hydrogen and oxygen in the escaping mixture. Defining the reservoir number ratio at the wind base as

\begin{equation}
    f_\mathrm{O} \equiv \left(\frac{n_\mathrm{O}}{n_\mathrm{H}}\right)_{\rm reservoir},
\end{equation}

we solve for an oxygen entrainment (fractionation) factor $x_\mathrm{O}\in[0,1]$ such that the escaping oxygen number flux is

\begin{equation}
    \phi_\mathrm{O} = \phi_\mathrm{H}\, f_\mathrm{O}\, x_\mathrm{O}.
\end{equation}

Physically, $x_\mathrm{O}$ encodes the competition between gravitational settling and diffusive drag by hydrogen (Appendix~\ref{app:fractionation_eqs}). When gravitational settling dominates, oxygen is inefficiently dragged and $x_\mathrm{O}\ll 1$; when drag dominates, oxygen is efficiently entrained and $x_\mathrm{O}\rightarrow 1$.

The hydrogen number flux $\phi_\mathrm{H}$ is set by mass-flux conservation. Writing the total mass flux in terms of number fluxes,

\begin{equation}
    F_\mathrm{mass} = m_\mathrm{H}\phi_\mathrm{H} + m_\mathrm{O}\phi_\mathrm{O},
\end{equation}

and substituting $\phi_\mathrm{O}=\phi_\mathrm{H} f_\mathrm{O} x_\mathrm{O}$ gives

\begin{equation}
    \phi_\mathrm{H} = \frac{F_\mathrm{mass}}{m_\mathrm{H} + m_\mathrm{O} f_\mathrm{O} x_\mathrm{O}},
\end{equation}

Because $x_\mathrm{O}$ depends on $\phi_\mathrm{H}$ through the diffusion--drag relations, we solve these equations by fixed-point iteration on $x_\mathrm{O}$ (or equivalently on $\phi_\mathrm{O}$), which converges rapidly for the regimes explored here.

Finally, we compute the mean molecular weight of the escaping H--O mixture (in hydrogen-mass units) as

\begin{equation}
    \mu_\mathrm{outflow} = \frac{\phi_\mathrm{H} m_\mathrm{H} + \phi_\mathrm{O} m_\mathrm{O}}{m_\mathrm{H}(\phi_\mathrm{H}+\phi_\mathrm{O})}.
\end{equation}

This $\mu_\mathrm{outflow}$ is fed back into the hydrodynamic solution in the coupled iteration described next.

\section{Numerical Solution for Coupled Mass-loss and Fractionation} \label{sec:self_consistent_fractionation}
We solve the coupled hydrodynamic mass-loss and hydrogen–oxygen fractionation problem using a self-consistent iterative scheme. While some recent population-level studies now couple mass-loss and multi-species fractionation (e.g., \citet{cherubim_oxidation_2025}), earlier approaches typically treated the two independently (e.g., \citet{luger_extreme_2015}). In our framework, the fractionated composition modifies the mean molecular weight $\mu_\mathrm{outflow}$ of the escaping gas and also modifies the XUV absorption through the mass absorption coefficient $\chi_\mathrm{XUV}$ evaluated for an atomic mixture at $R_\mathrm{XUV}$. We therefore iterate hydrodynamics and fractionation jointly until convergence.

For a single planet, the coupled iteration proceeds as follows:

\begin{enumerate}
    \item Initialize the outflow mean molecular weight $\mu_\mathrm{outflow}$ from the fully dissociated reservoir composition (Section~\ref{sec:mmw}), and construct an initial atomic mixture at $R_\mathrm{XUV}$ for evaluating $\chi_\mathrm{XUV}$.
    \item For the current $(\mu_\mathrm{outflow}, \chi_\mathrm{XUV})$, solve the hydrodynamic escape problem to obtain $(\dot{M}, c_s, R_s, R_\mathrm{XUV})$ (Section~\ref{sec:numerics}). We first compute the energy-limited solution and then, if the time-scale criterion indicates rapid recombination, adopt a recombination-limited solution (Section~\ref{sec:escape_rate}).
    \item Compute the outflow temperature used in the fractionation step. In the recombination-limited regime we set $T_\mathrm{outflow}=10^4\,\mathrm{K}$, consistent with the recombination thermostat; otherwise we map $T_\mathrm{outflow}$ from $(c_s,\mu_\mathrm{outflow})$ via $T_\mathrm{outflow}=c_s^2 m_\mathrm{H}\mu_\mathrm{outflow}/k_\mathrm{B}$.
    \item Using $(R_\mathrm{XUV}, \, T_\mathrm{outflow}, \, \dot{M})$, compute the total mass flux $F_\mathrm{mass}$ and apply the fractionation solver to obtain the number fluxes $\phi_s$ and entrainment factors $x_s$ (Section~\ref{sec:frac}).
    \item Update the atomic mixture at $R_\mathrm{XUV}$ from the escaping fluxes, $y_s \equiv \phi_s/\sum_{s}\phi_s$, and recompute $\mu_\mathrm{outflow}$ from the flux-weighted mixture:
    
    \begin{equation}
        \mu_\mathrm{outflow} = \frac{\sum_s \phi_s\, m_s}{m_\mathrm{H}\sum_s \phi_s}.
    \end{equation}
    
    \item Check convergence of $\mu_\mathrm{outflow}$. The iteration terminates when $|\mu_{\mathrm{outflow,new}}-\mu_{\mathrm{outflow,old}}|/\mu_{\mathrm{outflow,old}} < 10^{-5}$. Otherwise, return to step 2.
\end{enumerate}

A schematic overview of these steps is provided in the Appendix~\ref{app:figures}, Figure~\ref{fig:flowchart}.

\section{Model Parameters and Assumptions}
Here, we summarize the key physical parameters, opacities, and coefficients adopted throughout our framework, as well as the primary assumptions under our numerical implementation. These choices form the foundation for all hydrodynamic and fractionation calculations.

\subsection{Recombination, Diffusion, and Opacity Coefficients} \label{sec:recomb_opacities}
The recombination coefficient, governing the balance of ionization and recombination in the outflow region, is set to $\alpha_\mathrm{rec} = 2.6 \times 10^{-13} \, \mathrm{cm}^3\mathrm{/s}$ \citep{Storey1995-uv, osterbrock_2006}. 

The absorption of XUV photons is parameterized by atomic photoabsorption cross-sections $\sigma_{\mathrm{XUV,H}}$ and $\sigma_{\mathrm{XUV,O}}$, which we combine into a mass absorption coefficient $\chi_\mathrm{XUV}$ (units $\mathrm{cm^2\,g^{-1}}$) at $R_\mathrm{XUV}$. We adopt $\sigma_\mathrm{XUV} \approx 1.89 \times 10^{-18} \,\mathrm{cm}^2$ for both atomic hydrogen \citep{murray_atmospheric_2009} and oxygen, given their nearly identical photoionization cross-sections \citep{Verner1996-tw}, assuming a monochromatic XUV spectrum at $h\nu = 20$ eV. For an atomic H--O mixture with number fractions $y_\mathrm{H}$ and $y_\mathrm{O}$, we compute

\begin{equation}
    \chi_\mathrm{XUV} = \frac{y_\mathrm{H}\sigma_{\mathrm{XUV,H}} + y_\mathrm{O}\sigma_{\mathrm{XUV,O}}}{\mu_\mathrm{outflow} m_\mathrm{H}},
\end{equation}

where $\mu_\mathrm{outflow}$ is the mean particle mass in units of $m_\mathrm{H}$. In the coupled iteration, $(y_\mathrm{H},y_\mathrm{O})$ are updated from the escaping fluxes via $y_s=\phi_s/(\phi_\mathrm{H}+\phi_\mathrm{O})$. We note that this formulation implicitly assumes neutral absorbers at the wind base; in recombination-limited cases hydrogen may be partly ionized, which we neglect in the present work.

The opacity to outgoing thermal radiation, which characterizes radiative transfer through the escaping atmosphere, is taken as $\kappa_\mathrm{H/He} \approx 0.01 \, \mathrm{cm^2 g^{-1}}$ for a pure H/He atmosphere \citep[e.g.,][]{seager_exoplanet_2010, owen_evaporation_2017}, and $\kappa_\mathrm{H2O} \approx 1 \, \mathrm{cm^2 g^{-1}}$ for a pure steam atmosphere \citep[e.g.,][]{guillot_radiative_2010, seager_exoplanet_2010}. For mixed cases, we use an approximation: we perform a weighted interpolation of the pure-component opacities, e.g., a mixture of 10\% H$_2$O by mass with background hydrogen yields $\kappa_\mathrm{H_2, \, H2O} \approx 0.11 \, \mathrm{cm^2 g^{-1}}$.

We also adopt the binary diffusion coefficient for oxygen in hydrogen, $b_i = 4.8 \times 10^{17}\,T_\mathrm{outflow}^{0.75}$ $\mathrm{cm}^{-1}\,\mathrm{s}^{-1}$, following Table 1 from \citet{zahnle_mass_1986}. This coefficient is used in the fractionation calculations and reflects the drag force exerted on oxygen atoms by a hydrogen background at temperature $T_\mathrm{outflow}$ (see Appendix~\ref{app:fractionation_eqs}). The temperature dependence of $b_i$ is set by the hydrodynamic outflow temperature produced by the escape solution. The physically realized range of $T_\mathrm{outflow}$ across our models is discussed in detail in Appendix~\ref{app:fractionation_eqs}.

\subsection{Mean Molecular Weight and Atmospheric Composition} \label{sec:mmw}
Below $R_\mathrm{XUV}$, we assume the atmosphere is well mixed and reflects its bulk molecular composition as shown in Figure~\ref{fig:model}. The bolometric mean molecular weight $\mu_{\rm bolo}$ is computed from the molecular reservoir mass fractions and is used to evaluate the isothermal sound speed in the radiative layer, $c_{s,\rm bolo}=\sqrt{k_\mathrm{B}T_\mathrm{eq}/(\mu_{\rm bolo} m_\mathrm{H})}$. In this work, the mean molecular weight ranges from 2.35 $\mathrm{g \, mol^{-1}}$ (pure primordial H/He gas) to 18 $\mathrm{g \, mol^{-1}}$ (pure H$_2$O steam atmosphere). 

Above $R_\mathrm{XUV}$, we assume full dissociation of the molecular reservoirs and treat the escaping gas as an atomic mixture. The outflow mean molecular weight $\mu_\mathrm{outflow}$ is defined as the mean particle mass in units of $m_\mathrm{H}$. For a pure steam atmosphere, dissociation yields two H atoms and one O atom per H$_2$O molecule, giving $\mu_\mathrm{outflow}=6$. In mixed H$_2$+H$_2$O envelopes, $\mu_\mathrm{outflow}$ lies between $\simeq 1$ (hydrogen-dominated) and 6 (steam-dominated), and is updated self-consistently from the escaping fluxes in the coupled iteration (Section~\ref{sec:self_consistent_fractionation}).

The mean molecular weight used in the interior-structure model (Section~\ref{sec:models_params}), $\mu_{\rm bolo}$, is held constant and represents the bulk composition of the deep envelope. In contrast, the mean molecular weight $\mu_{\rm outflow}$, which is updated within the hydrodynamic/fractionation loop, applies only to the escaping atmosphere above $R_{\mathrm{XUV}}$. This updated value is not fed back into the interior-structure calculation. As a result, the planetary radius $R_p$ is determined solely by the fixed deep-envelope composition provided by the interior model. Incorporating changes in envelope composition due to fractionation would only be required in an evolutionary framework.

\subsection{Planetary Models and Input Parameters} \label{sec:models_params}
We apply our coupled mass-loss and fractionation model to the following scenarios:
 
\begin{enumerate}
    \item Young super-Earths and sub-Neptunes, with the goal of determining changes in atmospheric properties over time.
    \item Evolved super-Earths and sub-Neptunes, with the goal of determining mass-loss rates and the fractionation efficiency for various planet properties.
\end{enumerate}

Planetary radii are computed using a self-consistent interior structure model after \citet{dorn2015can, dorn_generalized_2017}, with recent updates from \citet{luo_majority_2024}. The model assumes an iron-rich core, a rocky mantle, and, where applicable, a surface ocean or an extended volatile envelope. Deep interior profiles are assumed to be adiabatic below the radiative-convective boundary $R_{\mathrm{rcb}}$, while above it, the radiative region rapidly relaxes toward an isothermal profile and we approximate it at the equilibrium temperature $T_{\mathrm{eq}}$. The thermodynamic state of the convective interior is set by the intrinsic luminosity, following the thermal evolution prescriptions of \citet{mordasini_planetary_2020}.

For sub-Neptunes, we model planets with masses of $2$-$15\,M_\oplus$. We generate both inflated, young planets initialized at an age of $\sim$100 Myr, as well as more compact, thermally evolved planets. For the young cases, the planet intrinsic luminosity is calculated following the luminosity model of \citep{mordasini_planetary_2020} and is a function of planet mass, atmospheric mass fraction, and planet age. For evolved sub-Neptunes, we instead prescribe a fixed intrinsic luminosity of $L_{\rm int}=10^{21}\,\mathrm{erg\,s^{-1}}$, representative of late-time cooling states. These planets are assigned equilibrium temperatures of 300 K and 1{,}000 K to explore a range of irradiation environments, and envelope compositions including atmospheres (3\% by planet mass) made up of either pure H/He, or H$_2$ containing 10–90\% water by atmospheric mass. The computed radii for these planets range between:

\begin{itemize}
    \item Young sub-Neptunes: approx. $2.24-2.86\,R_\oplus$ at 300 K, and $2.77-3.54\,R_\oplus$ at 1{,}000 K.
    \item Evolved sub-Neptunes: approx. $1.35-2.68\,R_\oplus$ at 300 K, and $1.53-3.36\,R_\oplus$ at 1{,}000 K.
\end{itemize}

For super-Earths, we model rocky planets with masses of $1$–$10\,M_\oplus$, and we adopt a representative atmospheric (or volatile) mass fraction of 3\% by planet mass as a baseline case that represents the layer overlying the differentiated interior. This choice is literature-motivated \citep[e.g.,][]{rogers_most_2024}) and allows us to illustrate the behavior of the coupled model. We adopt equilibrium temperatures of 400 K and 2{,}000 K to represent cooler, thermally evolved planets and hotter, more strongly irradiated planets, respectively. The computed radii range between approximately $1.25$-$2.1\,R_\oplus$ for planets at 400 K, and $2.1$–$2.45\,R_\oplus$ for planets at 2,000 K.
\section{Results} \label{sec:results}

\subsection{Mass Loss and Global Oxygen Fractionation Trends} \label{sec:evolved}
In this section we present results for evolved super-Earths and sub-Neptunes, mapping how instantaneous mass loss rates and oxygen retention vary across planetary mass, atmospheric composition, and irradiation, and identifying where fractionation becomes important and where oxygen is efficiently retained.

\begin{figure*}
    \centering
    \includegraphics[width=0.8\textwidth]{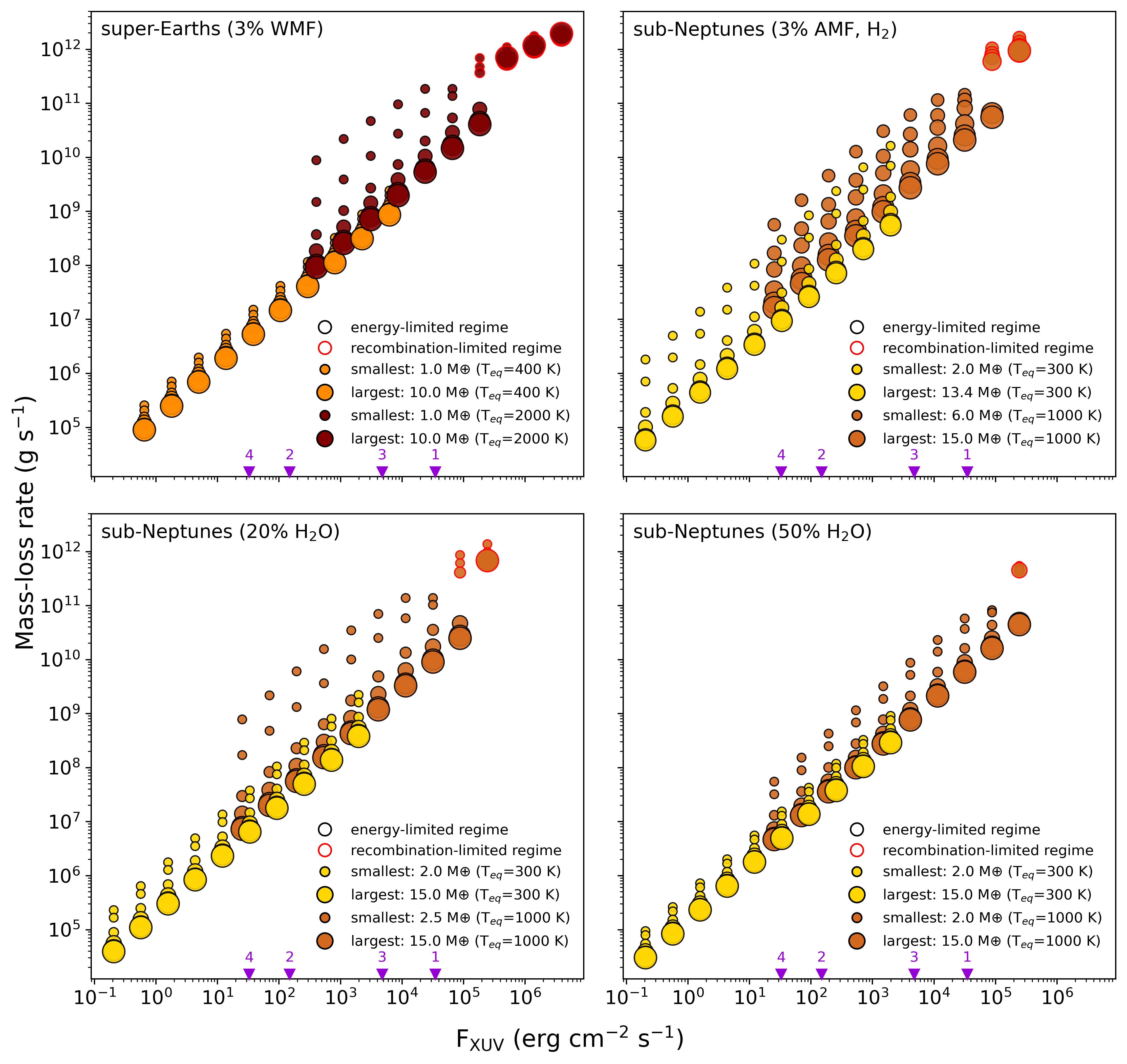}
    \caption{Mass loss rate $\dot{M}$ as a function of $F_\mathrm{XUV}$ for a range of planetary compositions and envelope structures. Each panel shows results for a different atmospheric configuration: (top left) super-Earths with 3\% WMF and a purely steam H$_2$O atmosphere; (top right) sub-Neptunes with 3\% atmospheric mass fraction (AMF) in H/He; (bottom left) sub-Neptunes with 3\% AMF in H$_2$ and 10\% envelope WMF; (bottom right) sub-Neptunes with 3\% AMF and 50\% envelope WMF. Each point corresponds to a specific model with varying $T_\mathrm{eq}$ and planet mass. Marker size reflects planetary mass, where the minimum, maximum, and few cases in between are plotted. Color denotes equilibrium temperature (from yellow to maroon with increasing $T_\mathrm{eq}$), and red point edges flag the recombination-limited regime. The purple indicators represent the XUV fluxes of certain scenarios and are there for context; indicators 1) and 2) correspond to the flux an Earth-like planet receives around a Sun-like star at an orbital period of 10 days, at 50 Myr and 5 Gyr, respectively; 3) and 4) are similar to 1) and 2) but for an M-star of mass 0.4$M_\odot$. Across the parameter space, higher XUV flux and lower mass lead to increased $\dot{M}$. The plateau of the $\dot M$ rates at high $F_{\mathrm{XUV}}$ is due to the physically-motivated imposed bounds on outflow temperature and sound speed.}
    \label{fig:Mdot_FXUV}
\end{figure*}

Figure \ref{fig:Mdot_FXUV} shows how the total atmospheric mass loss rates depends on incident XUV fluxes for different planetary masses, $T_\mathrm{eq}$, and envelope compositions. In general, (1) for a given $F_\mathrm{XUV}$, lower mass planets lose more atmospheric mass due to their less strong gravity; (2) for a given $F_\mathrm{XUV}$ and $M_{\mathrm p}$, planets with higher $T_\mathrm{eq}$ lose more atmospheric mass, due to their inflated radii; and (3) differences in atmospheric composition and envelope mass produce systematic offsets in the overall escape rates, especially for smaller mass planets. These offsets arise because higher $H_2O$ fractions may increase the outflow mean molecular weight and reduce the hydrogen number flux at the XUV base.

\begin{figure*}
    \centering
    \includegraphics[width=\textwidth]{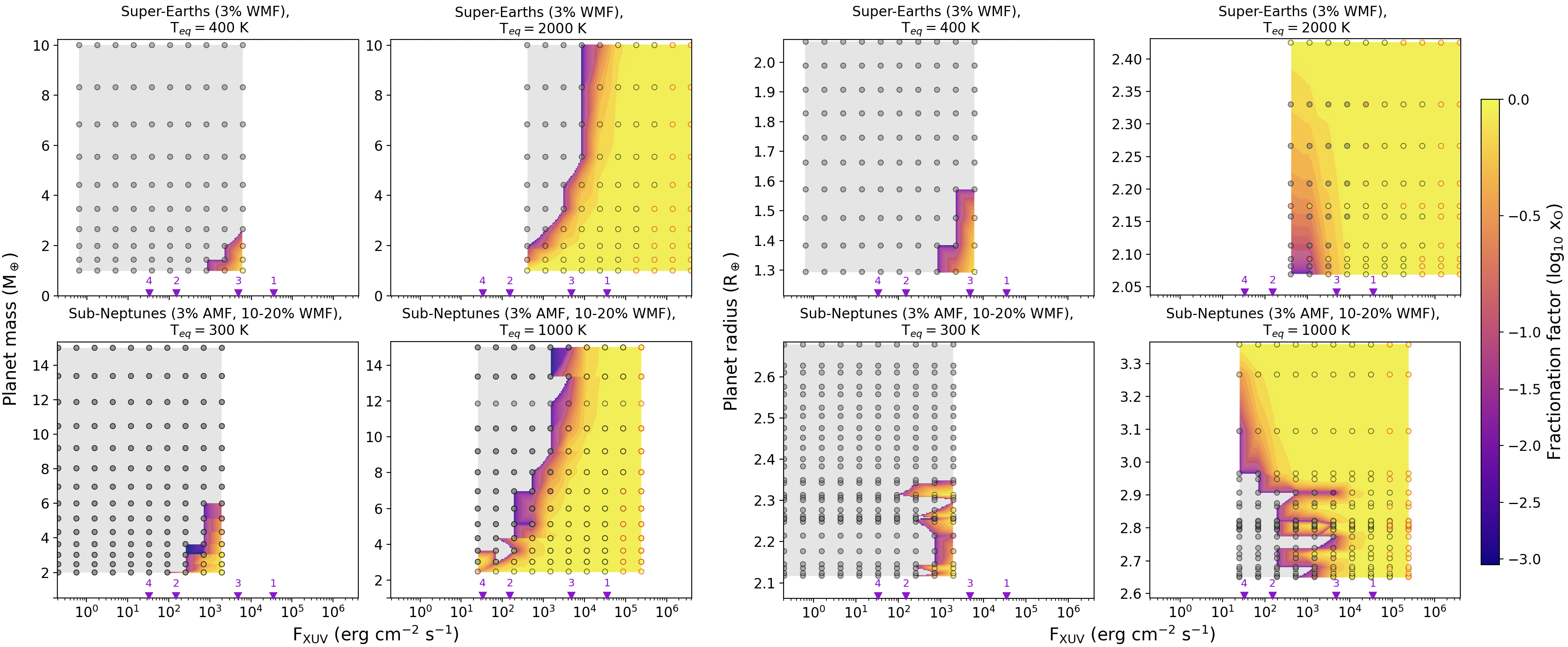}
    \caption{Oxygen fractionation factor map over planetary mass (left) and radius (right), and incident $F_\mathrm{XUV}$, pooled across envelope compositions and $T_\mathrm{eq}$. Gray shaded regions and scatter points show a fractionation factor for oxygen of 0 (no oxygen loss), while the colorbar shows increasing fractionation factors towards yellow colors. A fractionation factor of 1 means that oxygen is lost according to the mixing ratio (e.g., 1 oxygen atom per 2 hydrogen atoms in a pure steam atmosphere). Red point edges flag the recombination-limited regime. The purple indicators represent the XUV fluxes of certain scenarios and are identical to the indicators of Figure~\ref{fig:Mdot_FXUV}.}
    \label{fig:MR_FXUV}
\end{figure*}

It is clear that $F_\mathrm{XUV}$ is the main driving factor for total mass loss, and also determines whether oxygen loss is significant or suppressed, illustrated in Figure~\ref{fig:MR_FXUV} for both super-Earths and evolved sub-Neptunes. Oxygen loss is shown as the instantaneous oxygen fractionation factor $x_O$, ranging from 0 (no oxygen loss) to 1 (no fractionation). It directly relates to the escape flux $\phi_O$, as shown in Appendix~\ref{app:fractionation_eqs}, Eq. \ref{eq:xO_def}. Increasing $F_\mathrm{XUV}$ leads to a sharp transition between two regimes: one in which no oxygen is dragged along ($x_O$ = 0, gray regions) and another in which (close to) the bulk composition is lost ($x_O \approx 1$, yellow regions). Compared to the influence of $F_\mathrm{XUV}$, planetary mass and radius have secondary effects. For visual clarity, the sub-Neptune panels show only models with atmospheric WMF of 10 and 20\%, while cases with higher WMF (up to 90\%) are omitted from the figure but discussed in the text. Because planetary masses are fixed across these models, different compositions overlap exactly in mass space and appear superimposed, while in radius space the same combination produces a high density of points that can partially obscure the interpolated color gradient. The color map should therefore be interpreted as a qualitative guide to the fractionation regime, whereas the discrete points represent the physically meaningful sampling of parameter space. A closer look at Figure \ref{fig:MR_FXUV} shows that:

\begin{enumerate}
    \item Super-Earths (top panels) at $T_\mathrm{eq} = 2{,}000$ K span both energy-limited and recombination-limited escape, with the transition between regimes depending on $F_\mathrm{XUV}$. High oxygen loss in the mass regime may occur for $F_\mathrm{XUV} \gtrsim 5 \times 10^3 \, \mathrm{erg \, cm^{-2}\, s^{-1}}$, regardless of planet mass, but more massive planets ($M_p \gtrsim 4 \, M_\oplus$) lose significant amounts of oxygen ($x_O \ge 0.1$) for $F_\mathrm{XUV} \gtrsim 10^4 \, \mathrm{erg \, cm^{-2}\, s^{-1}}$ . Under the high $T_\mathrm{eq}$ and irradiation environment of these super-Earths, the entire space of modeled radii correlates with efficient oxygen loss. For the low end of XUV fluxes, smaller planets lose less oxygen (color gradient) or no oxygen (gray points). Cooler super-Earths at $T_\mathrm{eq}=400$ K remain in the energy-limited regime with negligible oxygen escape, with the exception of a narrow low mass-high irradiation regime, confirming that $F_\mathrm{XUV}$ is the main controlling parameter for oxygen loss. 
    \item Sub-Neptunes (bottom panels) at $T_\mathrm{eq} = 1{,}000$ K, similarly to the warm super-Earths, require high $F_\mathrm{XUV}$, both in the mass and radius regimes, to enable oxygen escape. Significant oxygen loss appears for $F_\mathrm{XUV} \gtrsim 10^3$ for most masses and radii, with more extreme thresholds ($F_\mathrm{XUV} \gtrsim 10^4$) needed for planets with masses over 10 $M_\oplus$. At masses below $\sim 6 \, M_\oplus$, lower $F_\mathrm{XUV}$ may drive oxygen loss and escape occurs under energy-limited conditions. The most inflated warm sub-Neptunes show oxygen loss under all irradiation environments. Similar to super-Earths, cooler sub-Neptunes at $T_\mathrm{eq} = 300$ K show no significant oxygen loss, regardless of radius or mass, reaffirming that high XUV flux is a prerequisite for fractionation to matter. It is important to note that for this category of planets, significant oxygen loss is likely to coincide with the epoch when these planets lose the bulk of their envelopes.
\end{enumerate}

Overall, we find that oxygen escape is suppressed or negligible across a large portion of the parameter space of interest. Only specific conditions of lower masses, larger radius, and sufficient $F_\mathrm{XUV}$ allow significant oxygen loss/bulk atmosphere loss, and these often require planets to be near or within the recombination-limited regime. Specifically, we find that in $\sim$ 75\% of our parameter space, $x_O=0$.

\begin{figure}
    \centering
    \includegraphics[width=\linewidth]{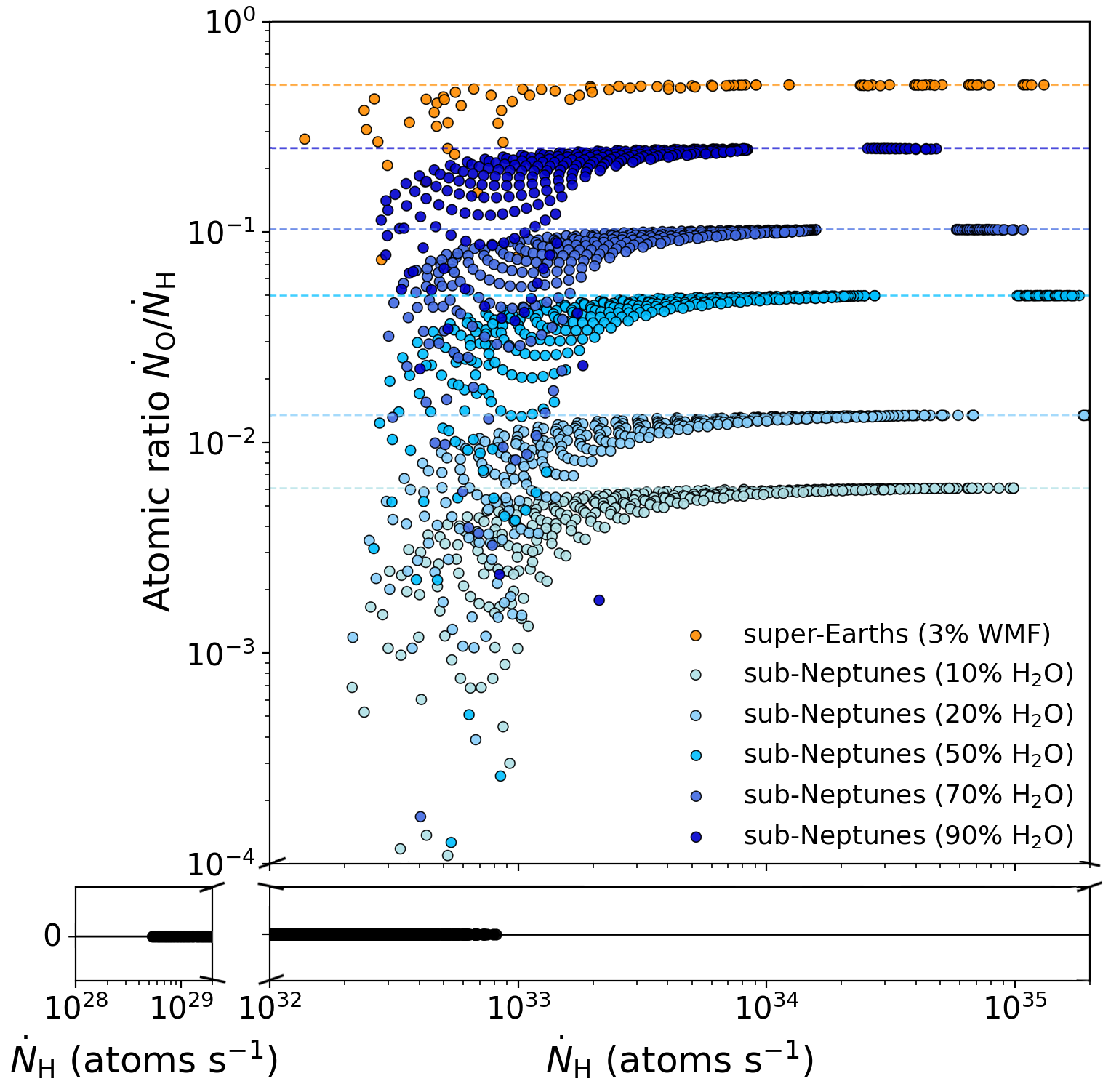}
    \caption{Atomic escape-flux ratio $\dot{N}_O/\dot{N}_H$ for all simulated planets spanning masses $1 - 15 M_\oplus$, equilibrium temperatures $300-2{,}000$ K, and a wide range of incident $F_\mathrm{XUV}$. A denser $F_\mathrm{XUV}$ sampling was done for the data shown in this plot compared to the previous figures, for better visualization. Cases where oxygen is lost ($x_O \neq 0$) are colored by planetary type and composition as indicated. Dashed lines show the mixing-limited maximum O/H ratios corresponding to each bulk water mass fraction, representing the upper limit for oxygen escape if it occurred at the bulk atmospheric abundance with a $x_O=1$. Black points at the 0 escape flux ratio line correspond to cases where oxygen is not lost, thus the ratio drops to 0. These cases make up $\sim$ 75\% of the data, overlap with each other, and span approximately 5 orders of magnitude of hydrogen loss (atoms s$^-{1}$).}
    \label{fig:oxygen_hydrogen_escape}
\end{figure}

To further examine oxygen’s contribution to the total escape flow, a direct diagnostic is the atomic ratio O/H escape flux. Figure~\ref{fig:oxygen_hydrogen_escape} presents this ratio for all simulated planets. Dashed lines in the figure indicate the mixing-limited maximum O/H ratio corresponding to each atmosphere composition (10-90\% water for sub-Neptunes, 100\% water for super-Earths, with, e.g., a O/H ratio of 0.5 for pure H$_2$O), representing the upper limit where no fractionation happens and oxygen escapes at the same rate as its atmospheric abundance. Calculated escape flux ratios $\dot{N}_O/\dot{N}_H$ are always lower than the mixing-limited ratios by construction, and lower escape flux ratios imply stronger fractionation. Values of $\dot{N}_O/\dot{N}_H=0$ correspond to cases where no oxygen is lost, with $x_O=0$. For cases where $x_O$ approaches unity, the escape flux ratios closely approach the mixing-limited ratios. We find that in cases where oxygen loss does occur ($\sim$25\% of our parameter space), hydrogen may escape in the same order of magnitude as oxygen, but more typically escapes one to two orders of magnitude more efficiently than oxygen in number flux, with extreme cases reaching $\dot{N}_O/\dot{N}_H=10^{-4}$. Lastly, consistent with H-dominated winds, planets with mixed H$_2$ + H$_2$O atmospheres exhibit systematically lower O/H ratios at a given hydrogen flux compared to pure steam super-Earths, because oxygen is always a trace component in a primarily hydrogen wind. This enhanced diffusive separation reduces drag on O atoms and further suppresses their escape.

\subsection{Atmosphere evolution over time} \label{sec:evolution}
In this second part, we apply our model to super-Earths and sub-Neptunes to determine how mass loss and fractionation alter their atmospheric properties over time. We focus on the possible water depletion for super-Earths and atmospheric enrichment of young sub-Neptunes over a time span of 200 Myrs; this corresponds to the timespan after disk dispersal when XUV-driven mass loss is expected to be strongest (Figure~\ref{fig:LXUV_evolution}). We note, however, that the duration of this high-activity phase can vary by nearly an order of magnitude depending on stellar type, and may extend to gigayear timescales for fully convective mid-to-late M dwarfs \citep{Pass_2023}.

While our model is designed to evaluate atmospheric escape and fractionation at fixed planetary snapshots, these instantaneous rates can offer valuable insight into the cumulative volatile loss experienced by a planet over time. To assess the broader implications of our results, we estimate the total water mass loss over 200 Myrs. Our estimates should be interpreted as upper bounds on potential water loss rather than precise evolutionary tracks. This is because we hold planetary radii fixed, whereas thermal contraction and envelope cooling over time would reduce $R_p$, requiring fully coupled interior---escape-evolution to also study the evolution of the planetary interior, which is beyond the scope of this work. A more complete treatment of water loss across time would require coupling radius evolution, thermal contraction, and stellar activity history, all of which would lower our estimates.

\subsubsection{How much water can be lost from Super-Earths due to oxygen fractionation?}
In order to estimate the maximum amount of water that can be lost over 200 Myrs from super-Earths with steam-dominated atmospheres (3\% WMF), we convert oxygen escape fluxes to the cumulative water loss. In practice, each escaping oxygen atom is paired with two hydrogen atoms, consistent with H$_2$O stoichiometry. This yields the rate of water molecules lost over time, which we multiply by 200 Myr to obtain the total water loss in Earth oceans.

\begin{figure}[ht]
    \centering
    \includegraphics[width=\linewidth]{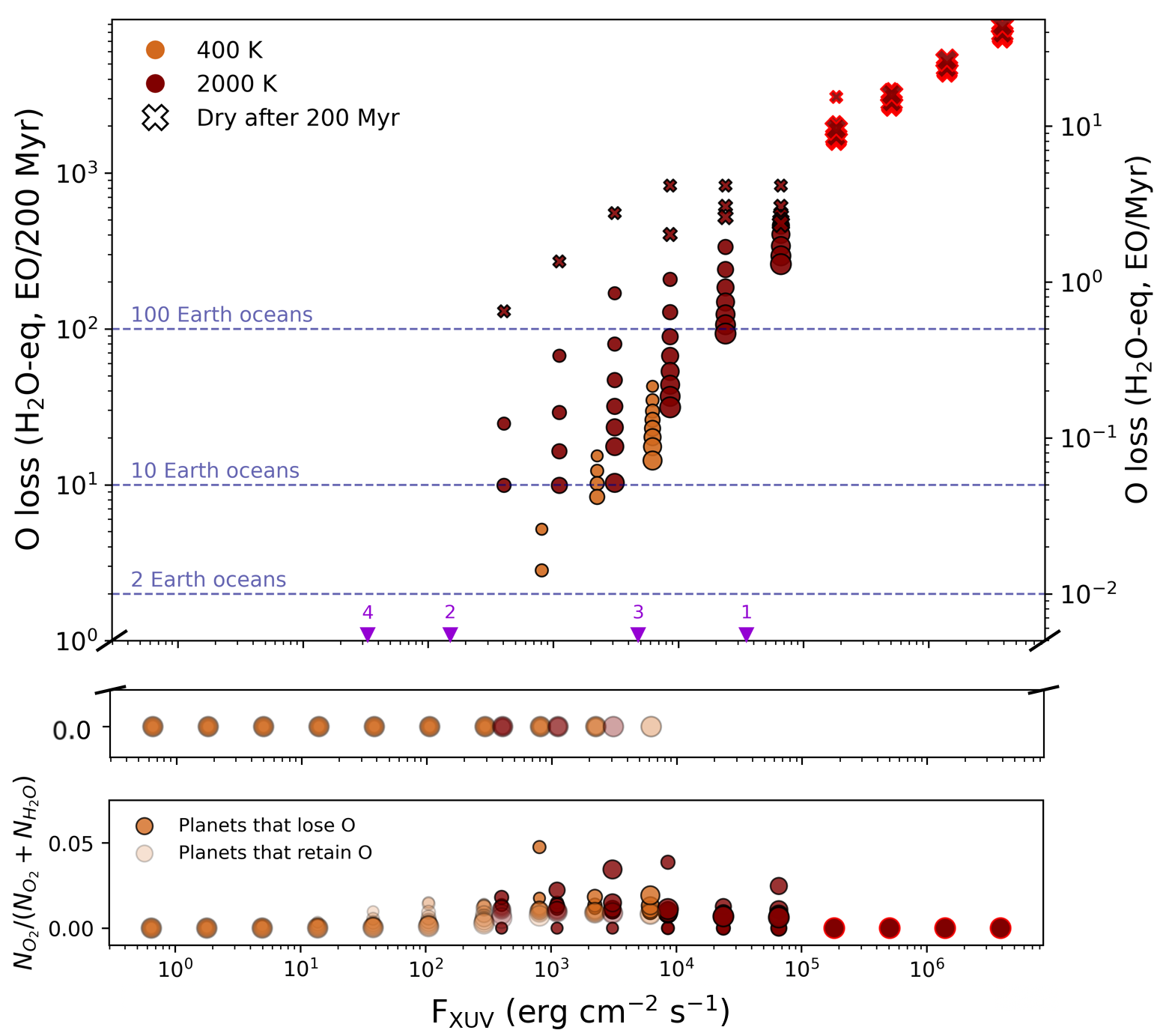}
    \caption{Estimated oxygen-equivalent water loss in Earth oceans for steam atmosphere (3\% WMF) super-Earths over 200 Myr (left y-axis) and 1 Myr (right y-axis) as a function of incident XUV flux, assuming constant escape rates over these timespans. Top panel: cases where water is lost due to oxygen loss; middle panel: cases where water is not lost due to oxygen being retained; bottom panel: resulting atmospheric oxygen content after recombination, i.e., oxygen enrichment after mass loss; opaque markers correspond to oxygen-loss cases (top panel), while faint markers correspond to oxygen-retention cases (middle panel). Marker size scales with planetary mass. Red edges indicate planets in the recombination-limited regime. The purple indicators represent the XUV fluxes of certain scenarios and are identical to the indicators of Figure~\ref{fig:Mdot_FXUV}.}
    \label{fig:water_loss_se}
\end{figure}

Figure~\ref{fig:water_loss_se} shows the cumulative oxygen loss, expressed in equivalent Earth oceans (EO) of H$_2$O as a function of incident XUV flux, as well as enrichment of oxygen in the remaining atmosphere due to preferential hydrogen loss in the bottom panel. This assumes instantaneous recombination of remaining hydrogen and oxygen into water and molecular oxygen. The results exhibit a threshold-dominated behavior controlled by whether oxygen is entrained in the escaping hydrogen flow, similar to Figure~\ref{fig:oxygen_hydrogen_escape}. For most planets at low irradiation levels ($F_{\rm XUV} \lesssim 10^{3}-10^{4}\,\mathrm{erg\,cm^{-2}\,s^{-1}}$), oxygen escape is suppressed ($\phi_{\rm O}=0$), resulting in negligible oxygen-equivalent-water loss despite ongoing hydrogen escape (middle panel). In this regime, oxygen is retained and oxygen-equivalent-water loss is effectively halted by diffusion-limited fractionation; here, we observe an atmospheric enrichment in the order of 4-5\% over the escape timescale (bottom panel).

At higher irradiation levels, a subset of planets transitions into a regime where oxygen becomes entrained and significant water loss occurs. For hot super-Earths ($T_{\rm eq}=2{,}000$ K, shown in dark red), this transition typically occurs at $F_{\rm XUV}\gtrsim 3 \cdot 10^{2}-10^{3} \, \mathrm{erg \, cm^{-2} \, s^{-1}}$, with oxygen loss increasing rapidly beyond this threshold. In these cases, cumulative losses of a few to hundreds of Earth oceans over 200 Myr are possible, and some planets are fully desiccated within this timescale (cross symbols). Cooler super-Earths ($T_{\rm eq}=400$ K) rarely enter this regime and generally retain their oxygen-equivalent-water inventories across the explored flux range.

These results imply that water loss during the evolution of super-Earths can remove up to the full 3\% water mass fraction assumed in our models for the lowest-mass, most irradiated planets. Since these cases result in complete depletion, our results provide a lower bound on the potential volatile loss: planets with higher initial water contents could plausibly lose even more. However, given that current constraints suggest most super-Earths contain $\lesssim 3$\% water by mass \citep{rogers_most_2024}, this level of loss may only be relevant for a small subset of the population. In reality, many super-Earths may be the remnants of sub-Neptune planets that have lost their primordial hydrogen envelopes due to intense stellar irradiation. The evolutionary pathway of these planets likely begins with envelopes that are not purely composed of water vapor but instead contain some amounts of hydrogen. Interestingly, \citep{werlen_sub-neptunes_2025} suggest that water-dominated envelopes are possible for super-Earths that formed within the ice-line. Such planets may be particularly susceptible to the levels of mass loss we estimate. Their bulk water content is typically limited to a few percent by mass, which is comparable in magnitude to the potential mass loss when the incident XUV fluxes are high.

\subsubsection{How much does atmospheric composition change with fractionated loss for sub-Neptunes?}
Atmospheric escape does not only remove mass; when the flow fractionates, it can also reshape the composition of the surviving envelope. For sub-Neptunes with mixed H$_2$–H$_2$O envelopes, preferential hydrogen loss tends to increase the envelope O/H ratio and therefore its effective metallicity. Here we estimate how efficiently this process can enrich young, warm sub-Neptunes during the post-disk epoch when stellar XUV emission is high.

We consider sub-Neptunes with envelope water mass fraction $Z_{\rm init}=0.2$ and $T_{\rm eq} = 1{,}000$ K. At each time step we evaluate the coupled hydrodynamic escape and H–O fractionation solution (Sections~\ref{sec:escape_model}–\ref{sec:self_consistent_fractionation}), and update the remaining hydrogen and oxygen reservoirs accordingly. To express the compositional outcome in an intuitive way, we assume that the retained H and O recombine into H$_2$O (and, when hydrogen becomes limiting, into O-bearing residual gas), and define the instantaneous envelope metallicity as the post-escape water mass fraction in the remaining envelope,

\begin{equation}
    Z(t) = \frac{M_{\rm H_2O}(t)}{M_{\rm env}(t)},
\end{equation}

where $M_{\rm env}$ is the total surviving envelope mass. This procedure is not a full evolutionary model because we hold the planetary radius fixed; as a result, the inferred enrichment represents an upper bound for our used luminosity, and for Sun-like hosts where contraction would typically reduce late-time escape rates, while it may be less conservative for M dwarfs with extended high-XUV phases \citep{Peacock2020-bs, Pass_2023}.

\begin{figure}
    \centering  
    \includegraphics[width=\linewidth]{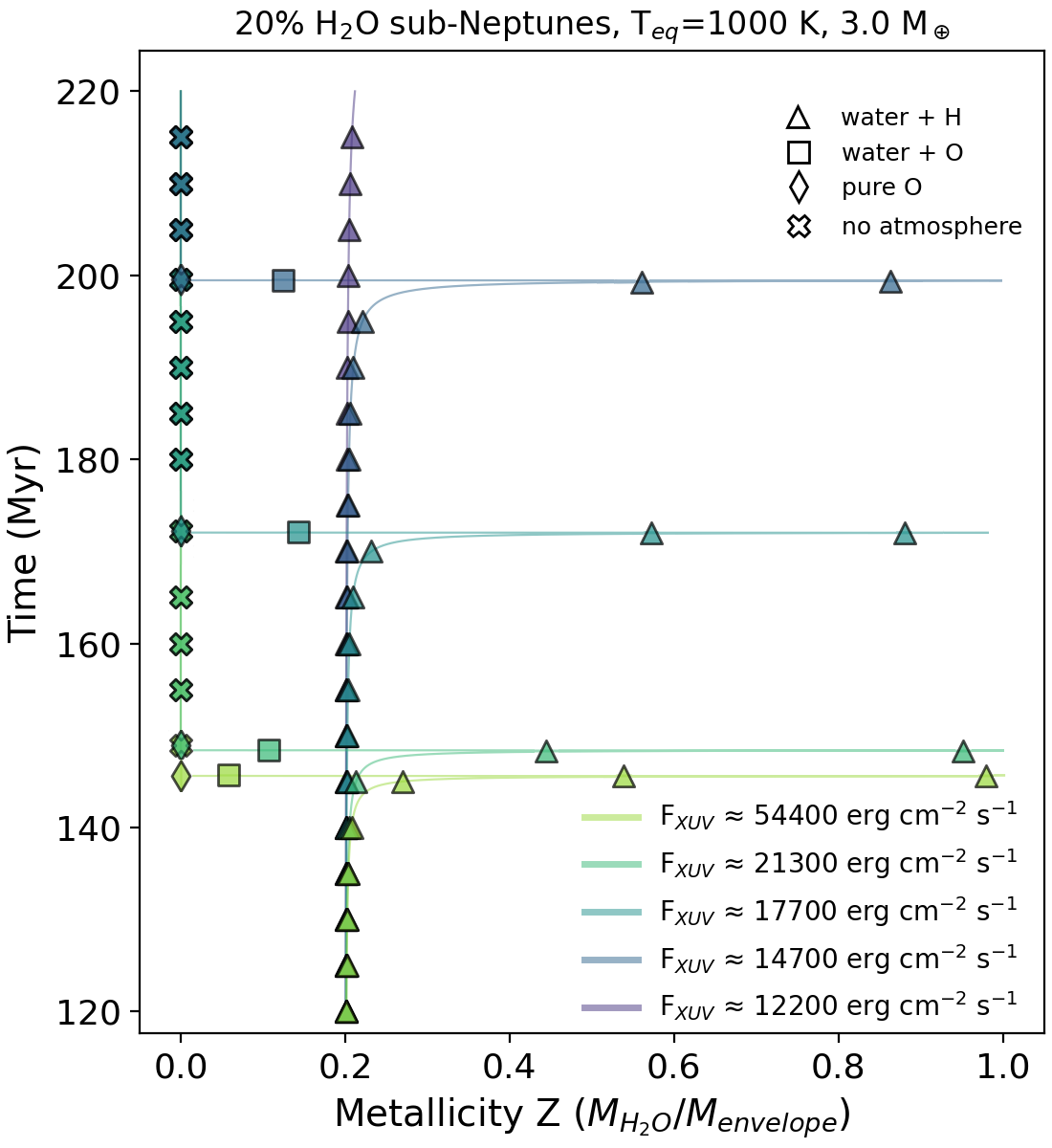}
    \caption{Evolution of the envelope metallicity $Z(t)$ for young sub-Neptunes with initial $Z=0.2$, $T_{\rm eq}=1{,}000$ K, and $M_p=3.0 \, M_\oplus$, shown as a function of incident $F_{\rm XUV}$, over time. Values approaching $Z_{\rm H_2O} \simeq 1$ correspond to steam-dominated surviving envelopes.}
    \label{fig:metallicity_time_FXUV}
\end{figure}

\begin{figure}
    \centering  
    \includegraphics[width=\linewidth]{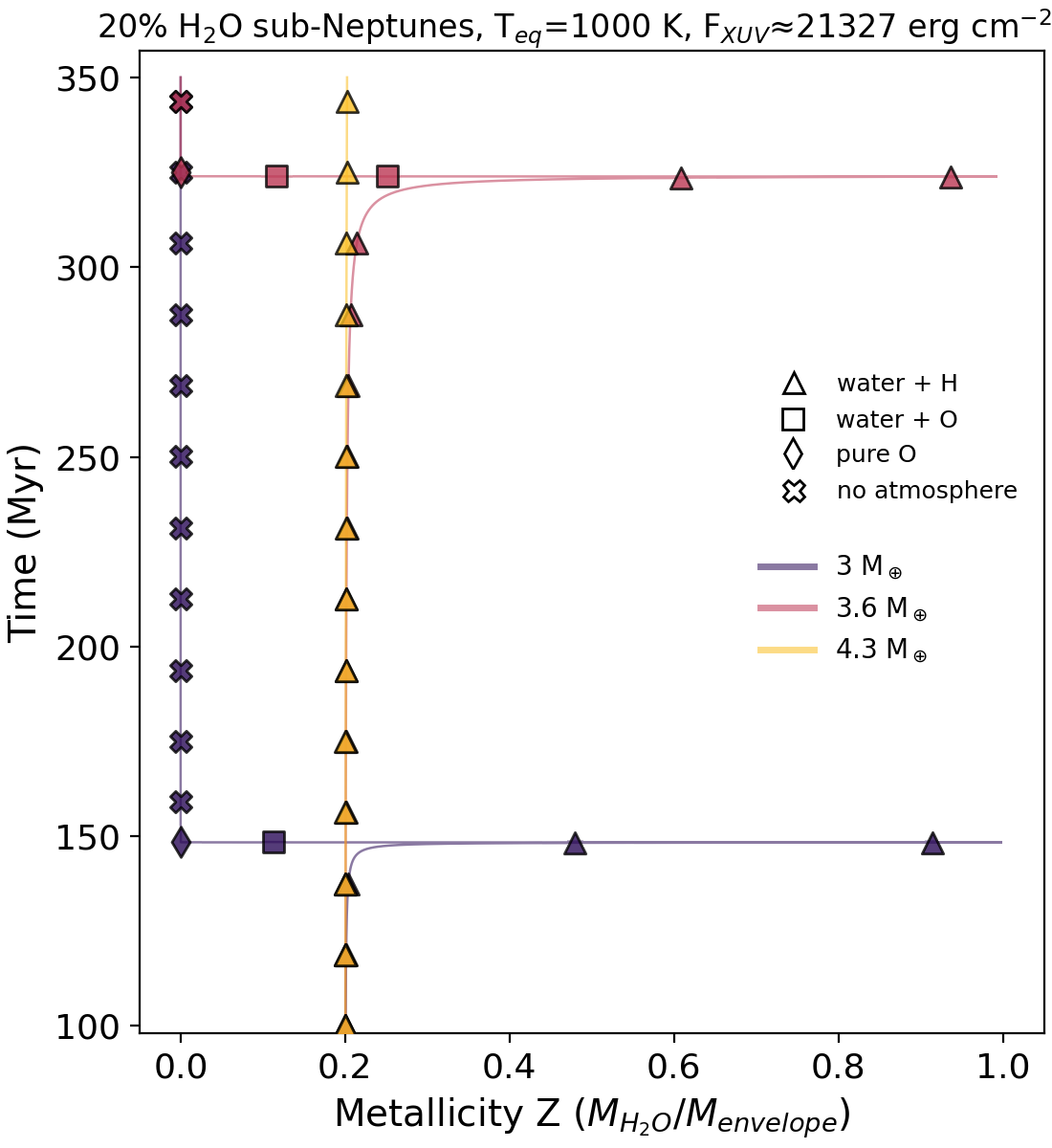}
    \caption{Evolution of the envelope metallicity $Z(t)$ for young sub-Neptunes with initial $Z=0.2$, $T_{\rm eq}=1{,}000$ K, and $F_{\rm XUV} = 21{,}327$, shown as a function of planetary mass, over time. Values approaching $Z_{\rm H_2O} \simeq 1$ correspond to steam-dominated surviving envelopes.}
    \label{fig:metallicity_time_mass}
\end{figure}

Figures~\ref{fig:metallicity_time_FXUV}–\ref{fig:metallicity_time_mass} illustrate that compositional evolution is strongly threshold dominated. At low irradiation, oxygen is efficiently retained ($x_{\rm O}\rightarrow 0$), but escape rates are small, so the envelope composition changes only weakly over $\sim 200$ Myr despite strong microphysical fractionation. At sufficiently high irradiation, escape becomes rapid and oxygen entrainment increases toward $x_{\rm O}\sim1$, causing the outflow composition to approach the bulk mixture and reducing the degree of preferential hydrogen loss. Consequently, substantial enrichment requires a narrow intermediate regime: hydrogen escape must be vigorous enough to remove a significant fraction of the envelope, while oxygen must remain at least partially retained long enough for $Z$ to rise.

In the fixed-mass sequence (Fig.~\ref{fig:metallicity_time_FXUV}), enrichment remains negligible below a characteristic XUV threshold, while above it the atmosphere transitions rapidly through successive compositional stages. As hydrogen is removed, the envelope becomes progressively more water rich, approaching $Z\simeq1$ (a steam-dominated remainder). For the most strongly irradiated cases, continued escape then removes the remaining volatile reservoir on short timescales, so that high-$Z$ phases can be transient and may precede complete atmospheric depletion after a very short pure-oxygen atmosphere state. For a fixed-flux(Fig.~\ref{fig:metallicity_time_mass}), the same behavior appears as a strong mass dependence: lower-mass planets evolve through enrichment and depletion much faster, whereas higher-mass sub-Neptunes retain their envelopes longer and exhibit slower or negligible enrichment over 200 Myr.

Our calculations show that fractionated escape can produce substantial water enrichment in young, warm sub-Neptunes as they evolve toward super-Earths. Significant enrichment occurs only for planets that lose most of their primordial H/He envelopes, and it happens rapidly, on timescales of order $\sim$ 10 Myr. Because our estimates represent upper limits on escape efficiency, the enrichment phase is likely more prolonged in reality.

A natural observational signature of this process is a transitional regime populated by increasingly enriched planets as they get closer to the radius valley. This leads to a clear prediction: the bulk of the sub-Neptune population should retain H/He-dominated envelopes with low water mass fractions, as constrained by geochemical considerations \citep{werlen_sub-neptunes_2025}, whereas only planets close or within the radius valley can develop strongly enriched envelopes, potentially transitioning to steam-dominated atmospheres. GJ\,9827\,d (1.98~$R_\oplus$, $T_\mathrm{eq}\sim620$~K) provides a compelling example. Its location near the radius valley and its inferred high metallicity (x500 times solar, with mass ratio $O/H \approx 4$) are fully consistent with our model predictions for a partially hydrogen-depleted planet sculpted by XUV-driven mass loss \citep{piaulet-ghorayeb_jwstniriss_2024}. We note that alternative mechanisms---such as magma–atmosphere chemical interaction and differential solubility---have also been proposed to produce oxygen-rich, low-C/O atmospheres \citep[e.g.,][]{Kite_2021, Seo_2024}, and a combination of escape and interior---atmosphere processes may operate in practice.
\section{Simplifications and Caveats}
Several assumptions and simplifications are inherent in our model framework:

\begin{itemize}
    \item We assume sufficient XUV energy to fully dissociate H$_2$ and H$_2$O molecules into atomic hydrogen and/or oxygen, neglecting partial dissociation or altitude-dependent dissociation scenarios. This is left for future work but could modestly alter $c_s$ and other parameters, that affecting both the total mass loss rate as well as the fractionated fluxes of different species.
    \item On the $T_{\rm eq}=300$ K sub-Neptune cases: since we assume the radiative atmosphere is isothermal and well-mixed up to $R_{\rm XUV}$, we ignore the possibility of H$_2$O condensation and cold-trap formation near these temperatures. Such a cold trap would deplete upper-atmosphere water and further suppress oxygen-bearing escape; including condensation could therefore generally reduce oxygen loss \citep{Pierrehumbert2010-tf, wordsworth_water_2013}. Additionally, for this mechanism to play a meaningful role, escape rates would need to be relatively high, which is typically not the case for the cool sub-Neptunes we model.
    \item Accumulated oxygen in the upper atmosphere could potentially shield deeper atmospheric layers or introduce radiative cooling effects, altering escape rates over time \citep{nakayama_survival_2022, kawamura_reduced_2024, yoshida_suppression_2024, broussard_impact_2025}. We do not currently model this effect.
    \item We do not model recombination or ionization reactions (e.g., hydrogen and oxygen recombining into water) within the outflow region. Instead, we focus on instantaneous fractionation under the assumption of stable photodissociation conditions.
    \item We ignore helium fractionation; in practice, the model can be adapted to 3-species fractionation for H, O, and He, but this falls outside the scope of the current study. When looking at mixed atmospheres, we model H$_2$ mixed with H$_2$O.
    \item Our water loss estimates in Section~\ref{sec:evolution} are based on a constant extrapolation of instantaneous escape rates over 200 Myr. In reality, as hydrogen is depleted and the envelope becomes increasingly oxygen-rich, mass loss rates may decline due to rising mean molecular weight, reduced scale height, or even the onset of diffusion-limited escape. At the same time, volatile release from the deep interior may replenish the atmosphere as gas is lost to space, particularly hydrogen \citet{dorn2021hidden, cherubim_oxidation_2025, steinmeyer2026coupled}. Consequently, the net atmospheric evolution remains uncertain, and resolving it requires fully coupled interior–atmosphere–escape models within a self-consistent evolutionary framework.
    \item If hydrogen becomes a trace species in a heavy oxygen-rich atmosphere, its escape may become diffusion-limited: H atoms must diffuse upward through a heavier background before they can escape, throttling the hydrogen loss rate. We do not model this regime in this study. 
\end{itemize}
\section{Conclusions} \label{sec:conclusion}
This study aimed to investigate how hydrodynamic atmospheric escape driven by stellar XUV not only removes atmospheric mass from planets, but also alters the chemical composition of their atmospheres through fractionation. In particular, we focused on the differential escape of hydrogen and oxygen in steam and mixed-composition atmospheres of super-Earths and sub-Neptunes, and explored the implications for planetary evolution.

To achieve this, we developed an open-source, self-consistent, coupled atmospheric escape-fractionation model called \textsc{BOREAS}. Our framework built on hydrodynamic mass loss calculations that span both energy-limited and recombination-limited regimes and incorporate a mass-dependent escape prescription that tracks the partitioning of hydrogen and oxygen fluxes. We validated our results against benchmark planets with published mass loss estimates-- a detailed comparison to literature mass-loss estimates for well-characterized systems is provided in Appendix\ref{sec:comparison}---and applied the model to a broad parameter space spanning planet mass, radius, equilibrium temperature, age, and incident XUV flux. Overall, we demonstrated that atmospheric escape is not just a bulk erosion process, but also involves a fractionation process that can change planetary compositions over time. Understanding the coupled dynamics of mass loss and fractionation is important for interpreting the nature of exoplanets, especially those with atmospheres interpreted to be water-rich: they may have originated from initially water-poor, H/He-dominated compositions and become H-depleted and water-rich during their fractionated mass loss histories. This is particularly relevant for planets within or near the radius valley; GJ\,9827\,d and its possible inferred steam atmosphere \citep{piaulet-ghorayeb_jwstniriss_2024} is a perfect example. Future extensions of this work may incorporate time-dependent thermal evolution and coupled radius contraction to fully capture the interplay between atmospheric loss, composition, and structure across planetary lifetimes. 

Our key findings are as follows:

\begin{enumerate}
    \item \textit{Oxygen is typically retained.} Across most of our parameter space, hydrogen escapes more readily than oxygen. The oxygen fractionation factor $x_O=0$ in these cases, and remains $<1$ for most super-Earths and sub-Neptunes, implying negligible oxygen escape---except under the most extreme irradiation or low-gravity conditions where it approaches unity.
    \item \textit{Mass loss is composition-dependent.} For a given planet mass, pure-H/He atmospheres typically lose mass at significantly higher rates than pure-steam atmospheres, due to their lower mean molecular weight and larger scale heights. Models that neglect composition effects may overestimate escape rates. This is evident in nearly all of our modeled escape rates for the benchmarked planets, where results vary by orders of magnitude depending on the assumed atmospheric composition. In turn, this implies that observed mass-loss rates provide a powerful diagnostic of atmospheric composition, as different compositions produce an observable imprint on inferred loss rates \citep{rogers2026using}.
    \item \textit{Fractionated escape can enrich atmospheres in water.} Sub-Neptunes with modest water abundances (10\%–20\%, equivalent to initial metallicity Z = 0.1–0.2) can become water dominated, steam worlds, or oxygen worlds over 200 Myr due to preferential hydrogen loss, but only when oxygen is also entrained in the flow. Thus, this enrichment operates only over a limited evolutionary window, applying primarily to low-mass sub-Neptunes experiencing intense irradiation, where oxygen remains entrained in the outflow. We therefore predict that such compositional transformations are most relevant for planets near or within the radius valley, where a residual atmosphere is retained while the majority of hydrogen has been preferentially lost to space.
\end{enumerate}



\begin{acknowledgments}
C.D. acknowledges support from the Swiss National Science Foundation under grant TMSGI2\_211313 and the COPL project funding for Evolution and Diversity of Super-Earth Atmospheres. This work has been carried out within the framework of the NCCR PlanetS supported by the Swiss National Science Foundation under grant 51NF40\_205606. 
J.E.O is supported by a Royal Society University Research Fellowship. This project has received funding from the European Research Council (ERC) under the European Union’s Horizon 2020 research and innovation programme (Grant agreement No. 853022). 
We thank the anonymous reviewers for their insightful comments, which greatly helped to improve this study.
We acknowledge the use of large language models (LLMs), including ChatGPT, to improve the grammar, clarity, and readability of the manuscript.
\end{acknowledgments}

\software{The package developed for this project is publicly available under a custom license and can be found in \url{https://github.com/ExoInteriors/BOREAS}.}


\appendix

\section{Atmospheric Escape Model}

\subsection{Hydrodynamic Escape Equations} \label{app:hydro_eqs}
The atmospheric structure below $R_{\mathrm{XUV}}$ follows the layered profile described by \citet{owen_mapping_2023}. Deep in the envelope, below the radiative–convective boundary $R_{\mathrm{RCB}}$, the gas is convective and approximately adiabatic. Above $R_{\mathrm{RCB}}$ and up to the planet’s photospheric radius $R_p$, the atmosphere becomes radiative and is approximately isothermal at the equilibrium temperature $T_{\mathrm{eq}}$. This isothermal radiative layer extends from $R_p$ outward to $R_{\mathrm{XUV}}$ where XUV photons begin to be absorbed. Within this region, the flow remains subsonic, close to hydrostatic balance, so the density at $R_{\mathrm{XUV}}$ is given by the usual exponential profile

\begin{equation}
\label{app:rho_XUV}
    \rho_{\mathrm{XUV}} \;=\; \rho_{\text{photo}}\,\exp\!\Biggl[\frac{G M_{p}}{c_{s}^2}\Bigl(\frac{1}{R_{\mathrm{XUV}}} - \frac{1}{R_{p}}\Bigr)\Biggr],
\end{equation}

where $\rho_{\text{photo}}$ is determined from the balance between gravity and radiative opacity at the base (e.g. $\rho_{\text{photo}} \sim g/(\kappa_{IR}\,c_{s}^2)$), $G$ is the gravitational constant, $M_{p}$ the planet mass, $R_p$ is the planet (transit) radius, and $c_{s}$ the sound speed in this (near--)isothermal region.

Above $R_{\mathrm{XUV}}$, the atmosphere is further heated by XUV radiation and the flow accelerates into a supersonic wind. Here the velocity structure is determined by solving the isothermal Parker wind equations. In our implementation, the sonic point is defined as

\begin{equation}
    R_{s} \;=\; \frac{G M_{p}}{2\,c_{s}^2},
\end{equation}

and its relation to $R_{\mathrm{XUV}}$ dictates which branch of the Parker wind solution is used:

\begin{itemize}
    \item If $R_{s} \ge R_{\mathrm{XUV}}$, the standard Parker wind solution is used, corresponding to the canonical integration constant $C=-3$ \citep{Parker1958-wq}.
    \item If $R_{s} < R_{\mathrm{XUV}}$, a modified solution with a constant parameter is adopted to ensure consistency with the subsonic base. In this case, the constant $C$ is chosen such that the velocity at the XUV base satisfies $u(R_{\mathrm{XUV}})=c_s$, ensuring continuity with the hydrostatic atmosphere below. This yields
    
    \begin{equation}
        C = 1 - 4\ln\!\left(\frac{R_{\mathrm{XUV}}}{R_s}\right) - \frac{4R_s}{R_{\mathrm{XUV}}},
    \end{equation}
    
which reduces to the standard value $C=-3$ as $R_{\mathrm{XUV}}\rightarrow R_s$. Across the parameter space explored here, deviations from $C=-3$ are of order unity and occur only when the sonic point lies interior to the XUV absorption radius.
\end{itemize}

Once the velocity profile $u(r)$ is obtained over a radial grid spanning from $R_{\mathrm{XUV}}$ outward, the density shape implied by continuity is

\begin{equation}
    \rho_{\rm shape}(r) \;=\; \left(\frac{R_s}{r}\right)^2\left(\frac{c_s}{u(r)}\right),
\end{equation}

which is dimensionless up to an overall normalization. We determine the physical normalization by enforcing the condition that the XUV optical depth equals unity at $R_{\mathrm{XUV}}$. Writing the XUV optical depth as

\begin{equation}
    \tau(R_{\mathrm{XUV}}) \;=\; \int_{R_{\mathrm{XUV}}}^{\infty} \rho(r)\,\chi_{\rm XUV}\,dr,
\end{equation}

where $\chi_{\rm XUV}$ is the mass absorption coefficient (units $\mathrm{cm^2\,g^{-1}}$), we compute the corresponding geometric column

\begin{equation}
    \tau_{\rm geom} \;\equiv\; \int_{R_{\mathrm{XUV}}}^{\infty} \rho_{\rm shape}(r)\,dr,
\end{equation}

and set the base density to

\begin{equation}
\label{app:rho_norm}
    \rho_{\mathrm{XUV}} \;=\; \frac{1}{\chi_{\rm XUV}\,\tau_{\rm geom}}.
\end{equation}

This normalization is performed at each trial value of $(R_{\rm XUV},c_s)$ during the energy-limited root-finding procedure, so that the density profile entering the momentum balance and optical-depth condition is updated self-consistently at every iteration. The full density profile is then $\rho(r)=\rho_{\mathrm{XUV}}\,\rho_{\rm shape}(r)$. Finally, the mass-loss rate follows from

\begin{equation}
    \dot{M}=4\pi R_{\rm XUV}^2\,\rho_{\rm XUV}\,u(R_{\rm XUV}).
\end{equation}

\section{Atmospheric Mass Fractionation}

\subsection{Fractionation Factor Derivation} \label{app:fractionation_eqs}
We follow the diffusion--drag framework of \citet{zahnle_mass_1986, hunten_mass_1987}, which describes the entrainment of oxygen by an escaping hydrogen flow through binary diffusion, and adopt the formulation and notation used by \citet{odert_escape_2018} in the context of hydrodynamic escape. The oxygen entrainment (fractionation) factor $x_{\rm O}$ is defined by

\begin{equation}
\label{eq:xO_def}
    \phi_{\rm O} = \phi_{\rm H}\,f_{\rm O}\,x_{\rm O},
\end{equation}

where $f_{\rm O}=(n_{\rm O}/n_{\rm H})_{\rm reservoir}$ is the reservoir number ratio at the base of the wind, and $\phi_{\rm O}, \,\phi_{\rm H}$ are the escape number fluxes of oxygen and hydrogen, respectively. In the energy-limited (hydrodynamic) regime, the diffusion--drag solution yields

\begin{equation}
\label{eq:xO}
    x_{\rm O} = 1 - \frac{g\,(m_{\rm O}-m_{\rm H})\,b_{\rm HO}(T_{\rm outflow})}{k_{\rm B}T_{\rm outflow}\,\phi_{\rm H}\,(1+f_{\rm O})},
\end{equation}

where $g=GM_p/R_{\rm XUV}^2$ is the gravitational acceleration at the wind base, $b_{\rm HO}(T)$ is the binary diffusion coefficient for O in an H background (units $\mathrm{cm^{-1}\,s^{-1}}$), and $\phi_{\rm H}$ is the hydrogen \emph{number} flux (units $\mathrm{cm^{-2}\,s^{-1}}$). We enforce $x_{\rm O}\in[0,1]$; when Eq.~(\ref{eq:xO}) would predict $x_{\rm O}\le 0$, oxygen is not entrained and the escape becomes effectively diffusion-limited for oxygen.

The outflow temperature used in the diffusion coefficient is
\begin{equation}
    T_{\rm outflow}=
    \begin{cases}
    10^{4}\,{\rm K}, & \text{(RL)}\\[4pt]
    \dfrac{c_s^{2}\,m_{\rm H}\,\mu_{\rm outflow}}{k_{\rm B}}, & \text{(EL)}.
    \end{cases}
\end{equation}

In our framework, the outflow temperature $T_{\rm outflow}$ is not determined by radiative transfer, but follows directly from the hydrodynamic solution. In energy-limited cases, $T_{\rm outflow}$ is computed from the sound speed and the converged outflow mean molecular weight, while in recombination-limited escape it is fixed to $10^4$ K by the recombination thermostat. Across the parameter space where mass loss and fractionation are non-negligible, $T_{\rm outflow}$ typically lies between $(1\text{--}5) \cdot 10^3$ K for energy-limited escape, while recombination-limited cases reach $T_{\rm outflow} \sim 10^4$ K by construction. Only weakly irradiated planets with negligible escape exhibit lower temperatures, in the hundreds. As a result, the temperature dependence of the binary diffusion coefficient, $b\propto T_{\rm outflow}^{0.75}$, introduces only moderate variation in the coupling strength between hydrogen and oxygen across the relevant models.

\section{Comparison to estimated mass loss rates} \label{sec:comparison}
To validate and contextualize our model, we compare our predicted mass loss rates to previously simulated values for (well-) characterized exoplanet systems that span a range of atmospheric compositions, sizes, and irradiation levels. For this step, we compute escape and fractionation parameters based on published values of planetary mass, radius, $F_{\mathrm XUV}$, and $T_{\mathrm eq}$. We draw comparisons for the 2 extreme cases of atmospheric composition: pure H$_2$O atmosphere, and pure H/He envelope (approximately 30\% helium and 70\% hydrogen by mass). We show all results in Table~\ref{tab:mass_loss_rates}.

We begin with the TOI-431 system, which hosts at least three planets, with TOI-431 b and d being of particular interest. TOI-431 b is a short-period super-Earth, while TOI-431 d is a larger sub-Neptune. Observations by \citet{osborn_toi-431hip_2021} suggest that TOI-431 d likely retains a substantial volatile envelope, which could be either a light H/He layer ($\sim$ 3.6\% by mass) or a heavier water-rich layer ($\sim$ 33\% by mass). To bracket this uncertainty, we compute mass loss rates for both compositions to compare to previous calculations \citep{jiang_estimating_2025}, as seen in Table~\ref{tab:mass_loss_rates}. In the case of steam, we find that oxygen is strongly retained ($x_O = 0$), potentially enriching the atmosphere as hydrogen escapes. While the present-day mass loss rates for the two compositions of TOI-431 d appear comparable, their long-term evolutionary implications diverge significantly: With a H/He envelope, the planet is likely undergoing slow but continuous envelope erosion; in the steam case, the atmosphere is both more tightly bound and composed of heavier volatiles, with hydrogen escaping preferentially and nearly all oxygen being retained. This suggests that a water-rich planet could retain much of its volatile inventory over gigayear timescales, even under moderate XUV flux. As such, TOI-431 d exemplifies how $\dot{M}$ alone is insufficient to constrain atmospheric evolution without knowing the envelope's composition.

We also include the TRAPPIST-1 system, a benchmark for low-mass, rocky, cool exoplanets orbiting an M-dwarf star. We compare our results to the energy-limited escape rates compiled by \citet{becker_coupled_2020}, and recalculate loss rates assuming pure steam atmospheres for all planets. This setup provides insight into whether significant water depletion could occur in the absence of primordial hydrogen, particularly in early outgassed secondary atmospheres. We find that our escape rates are generally comparable to, and in several cases higher than, those reported by \citet{becker_coupled_2020}. While both approaches recover the same monotonic decrease in mass loss with orbital distance, the differences likely reflect our self-consistent treatment of the flow structure and XUV absorption radius, as well as the explicit inclusion of compositional effects. Importantly, escape remains strongly fractionated in all cases, with hydrogen being preferentially lost while oxygen is retained ($x_O = 0$). If water is the dominant outgassed volatile, our results suggest that TRAPPIST-1 planets could evolve toward oxygen-rich atmospheres, with the ultimate outcome depending on the efficiency of surface sinks in removing excess oxygen. While we observe the same trend as \citet{becker_coupled_2020} in decreasing XUV flux and mass loss rate with orbital distance, the systematically lower rates highlight the need for composition- and regime-aware models in assessing long-term atmospheric evolution for terrestrial planets.

Moreover, we apply our mass loss framework to three additional systems: K2-18 b, a sub-Neptune candidate, and HD 63433 c, a young sub-Neptune. For K2-18 b, we run our models both for a pure H/He case (no fractionation) and for a pure water composition (fully fractionated) to bracket the possible extremes. Our predicted escape rates are lower by approximately an order of magnitude, for both compositions, compared to the estimate from \citet{santos_high-energy_2020}, who applied an energy-limited model based on partial Lyman-$\alpha$ transit data. We attribute this difference to two factors: i) our self-consistent hydrodynamic and compositional treatment of the atmosphere in \textsc{BOREAS}, which explicitly solves for the wind structure, density normalization, and XUV absorption radius, rather than assuming a fixed energy-limited scaling, and ii) although our model yields an XUV absorption radius slightly larger than the planetary radius ($R_{\mathrm XUV}/R_p \simeq 1.1$), the resulting influence on the mass-loss rate is limited. We emphasize that the true escape rates are expected to lie between our two bracketing cases, given the range of possible atmospheric compositions of K2-18 b.

For HD 63433 c, we find that \textsc{BOREAS} predicts escape rates approximately an order of magnitude lower than the value reported by \citet{zhang_detection_2022}. who modeled the observed Ly$\alpha$ absorption using 3D hydrodynamics coupled to radiative transfer and nonequilibrium thermochemistry. While their model reproduces the overall blue-wing light curve, they note that it does not match the detailed velocity-dependent absorption and that the inferred signal is sensitive to assumptions about the stellar high-energy spectrum and the planet’s (currently unconstrained) mass. In our framework, a smaller effective XUV absorption radius, together with differences in the thermochemical and ionization structure of the outflow, leads to a reduced $\dot M$. Reconciling such lower escape rates with the observed Ly$\alpha$ depth would likely require differences in the neutral hydrogen distribution and/or wind–outflow interaction (e.g., processes that broaden the absorption in velocity space). Given the youth and activity of the host star, part of the observed helium signal may also arise from stellar variability. 

\begin{table*}
    \centering
    \caption{Mass loss rate comparison for different exoplanets. In our comparisons, we assume either a pure steam or pure H/He atmosphere. The $F_{\rm XUV}$ values and mass loss rates are static and correspond to present-day conditions; comparisons are intended to be illustrative rather than evolutionary.}
    \label{tab:mass_loss_rates}
    \begin{tabular}{lcccc}
            \toprule
            \shortstack{Planet} & 
            \shortstack{$F_\mathrm{XUV}$ \\ {[erg cm$^{-2}$ s$^{-1}$]}} & 
            \shortstack{Previous Mass \\ Loss rates {[g/s]}} & 
            \shortstack{This Study \\ (Fractionated \\ H$_2$O Atmosphere) \\ {[g/s]}} & 
            \shortstack{This Study \\ (H/He Envelope, \\ no fractionation) \\ {[g/s]}} \\
            \midrule
            TOI-431 b    & $70,286$\textsuperscript{[a]}  & $10^{10.51}$\textsuperscript{[a]}         & $10^{9.972}$          & No convergence \\
            TOI-431 d    & $93$\textsuperscript{[a]}      & $10^{9.140}$\textsuperscript{[a]}         & $10^{8.742}$          & $10^{9.122}$ \\
            \midrule
            TRAPPIST-1 b & $2{,}935$\textsuperscript{[b]} & $8.29 \cdot 10^{8}$\textsuperscript{[b]}  & $5.074 \cdot 10^{8}$  & -                 \\
            TRAPPIST-1 c & $1{,}565$\textsuperscript{[b]} & $3.62 \cdot 10^{8}$\textsuperscript{[b]}  & $2.663 \cdot 10^{8}$  & -                 \\
            TRAPPIST-1 d & $788$\textsuperscript{[b]}     & $2.61 \cdot 10^{8}$\textsuperscript{[b]}  & $1.731 \cdot 10^{8}$  & -                 \\
            TRAPPIST-1 e & $456$\textsuperscript{[b]}     & $9.08 \cdot 10^{7}$\textsuperscript{[b]}  & $8.799 \cdot 10^{7}$  & -                 \\
            TRAPPIST-1 f & $264$\textsuperscript{[b]}     & $6.58 \cdot 10^{7}$\textsuperscript{[b]}  & $4.767 \cdot 10^{7}$  & -                 \\
            TRAPPIST-1 g & $178$\textsuperscript{[b]}     & $4.78 \cdot 10^{7}$\textsuperscript{[b]}  & $3.140 \cdot 10^{7}$  & -                 \\
            TRAPPIST-1 h & $102$\textsuperscript{[b]}     & $2.90 \cdot 10^{7}$\textsuperscript{[b]}  & $2.370 \cdot 10^{7}$  & -                 \\
            \midrule
            K2-18 b      & $107.9$\textsuperscript{[c]}   & $3.50 \cdot 10^{8}$\textsuperscript{[c]}  & $2.522 \cdot 10^{7}$  & $3.279 \cdot 10^{7}$ \\
            \midrule                                                        
            HD 63433 c   & $6{,}000$                      & $2.1 \cdot 10^{10}$\textsuperscript{[d]}  & $1.249 \cdot 10^{9}$  & $1.860 \cdot 10^{9}$ \\
            \midrule
            TOI 560 b    & $5{,}000$\textsuperscript{[e]} & $1.61 \cdot 10^{11}$\textsuperscript{[e]} & $1.740 \cdot 10^{9}$  & $3.885 \cdot 10^{9}$ \\
            TOI 1430.01  & $6{,}800$\textsuperscript{[e]} & $1.32 \cdot 10^{11}$\textsuperscript{[e]} & $1.495 \cdot 10^{9}$  & $4.331 \cdot 10^{9}$ \\
            TOI 1683.01  & $12{,}000$\textsuperscript{[e]}& $2.48 \cdot 10^{10}$\textsuperscript{[e]} & $3.057 \cdot 10^{9}$  & $1.008 \cdot 10^{10}$ \\
            TOI 2076 b   & $6{,}000$\textsuperscript{[e]} & $2.43 \cdot 10^{10}$\textsuperscript{[e]} & $2.787 \cdot 10^{9}$  & $7.204 \cdot 10^{9}$ \\
            \bottomrule
        \end{tabular}

    \vspace{0.4em}
        \parbox{\textwidth}{\centering\footnotesize
        \textsuperscript{a} After \cite{jiang_estimating_2025}.\\
        \textsuperscript{b} After \cite{becker_coupled_2020}.\\
        \textsuperscript{c} After \cite{santos_high-energy_2020}.\\
        \textsuperscript{d} After \cite{zhang_detection_2022}.\\
        \textsuperscript{e} After \cite{Zhang2023-kc}.
    }
\end{table*}

Lastly, we benchmark our framework against the four young mini-Neptunes with metastable helium detections presented by \citet{Zhang2023-kc}: TOI-560 b, TOI-1430.01, TOI-1683.01, and TOI-2076 b. For these planets, the authors report XUV irradiation levels and infer present-day escape strengths via envelope-loss timescales. Converting their "time to lose 1\% of the planet mass" into mass-loss rates yields their estimated values shown in Table~\ref{tab:mass_loss_rates}. We show the values for their Parker-wind retrievals for direct comparison to our model. Using the same planetary parameters and irradiation levels, our model predicts differences of up to two orders of magnitude compared to \citet{Zhang2023-kc} inferred rates, depending on envelope composition. These systems provide a direct comparison point anchored in spectroscopic escape detections in the mini-Neptune regime. 
 
In general, pure steam atmospheres yield lower $\dot{M}$ values than equivalent pure H/He envelopes, due to their higher mean molecular weight and lower scale heights. Our results highlight the importance of composition-sensitive modeling frameworks for interpreting present-day escape and volatile retention. While our estimates differ quantitatively from literature values, these variations are expected due to differences in assumed opacity, heating efficiency, and hydrodynamic structure, as mentioned in the case of K2-18 b. Importantly, our results reinforce the need for composition-dependent models when interpreting atmospheric evolution across exoplanet populations.

\newpage

\section{Flowchart of the \textsc{BOREAS} framework} \label{app:figures}

\setcounter{figure}{8}
\renewcommand{\thefigure}{\arabic{figure}}
    
\begin{figure}[H]
    \centering
    \caption{Flowchart for the coupled hydrodynamic escape and H--O fractionation model. The outer loop iterates on $\mu_{\rm outflow}$ (and thus the atomic mixture used for $\chi_{\rm XUV}$) to obtain a self-consistent $R_{\rm XUV}$, $c_s$, $\dot{M}$, and $(\phi_{\rm H},\phi_{\rm O})$. EL: energy--limited; RL: recombination--limited.}
    \label{fig:flowchart}
    \begin{tikzpicture}[node distance=1.5cm, scale=0.7, transform shape]

    \node (start) [startstop] {Initialize $M_p$, $R_p$, $T_{eq}$, $F_{XUV}$; set initial dissociated mixture $(y_{\rm H},y_{\rm O})_{\rm init}$ and $\mu_{\rm outflow,init}$};
        
    \node (set_mu) [process, below of=start] {Set current $\mu_{\rm outflow}$ and compute $\chi_{\rm XUV}(y_{\rm H},y_{\rm O},\mu_{\rm outflow})$};
    
    \node (massloss_init) [process, below of=set_mu] {Hydrodynamic escape: set trial range for $R_{\rm XUV}$ and bracket $c_s$};
    \node (compute_el) [process, below of=massloss_init] {Compute analytic $\dot M_{EL}$};
    \node (guess_cs) [process, below of=compute_el] {Guess $c_s$ and compute $\dot M(c_s)$ via Parker--wind integration (uses $\chi_{\rm XUV}$)};
    \node (check_cs) [decision, below of=guess_cs, yshift=-0.5cm] {Does $\dot M(c_s)=\dot M_{EL}$?};
    \node (adjust_cs) [process, right of=check_cs, xshift=4cm] {Adjust $c_s$ guess};
    \node (update_reuv) [process, below of=check_cs, yshift=-1cm] {Update $R_{\rm XUV}$ by optical depth condition $\tau(R_{\rm XUV})=1$ (uses $\chi_{\rm XUV}$)};
    \node (massloss_done) [process, below of=update_reuv] {Converged hydro: $R_{\rm XUV}$, $c_s$, $\dot M$};

    \node (classify) [decision, below of=massloss_done, yshift=-0.5cm, text width=2.8cm] {Time--scale ratio $\mathcal{R}_t<1$?};
    \node (el_branch) [process, below of=classify, xshift=-3.5cm, yshift=-1cm] {Adopt EL solution};
    \node (rl_branch) [process, below of=classify, xshift=4cm, yshift=-1cm] {Adopt RL closure: set $c_s$ for $T\simeq 10^4$ K and re-solve for $R_{\rm XUV}$ and $\dot M$};
    \node (select_branch) [process, below of=classify, yshift=-2.5cm] {Select branch solution for $\dot M$, $R_{\rm XUV}$, $c_s$};

    \node (calc_tout) [process, below of=select_branch] {Set $T_{\rm outflow}$: if RL, $T_{\rm outflow}=10^4$ K; else $T_{\rm outflow}=c_s^{2} m_{\rm H}\mu_{\rm outflow}/k_{\rm B}$};
    \node (calc_fmass) [process, below of=calc_tout] {Compute $F_{\rm mass}=\dot M/(4\pi R_{\rm XUV}^2)$};
    \node (frac_solve) [process, below of=calc_fmass] {Solve H--O fractionation: obtain $x_{\rm O}$, $\phi_{\rm H}$, $\phi_{\rm O}$ with $F_{\rm mass}=m_{\rm H}\phi_{\rm H}+m_{\rm O}\phi_{\rm O}$};
    \node (update_mu) [process, below of=frac_solve] {Update $(y_{\rm H},y_{\rm O})=\phi/(\phi_{\rm H}+\phi_{\rm O})$ and $\mu_{\rm outflow,new}=\frac{\phi_{\rm H}m_{\rm H}+\phi_{\rm O}m_{\rm O}}{m_{\rm H}(\phi_{\rm H}+\phi_{\rm O})}$};

    \node (check_mu) [decision, below of=update_mu, yshift=-0.5cm] {Converged $\mu_{\rm outflow}$?};
    \node (iter_mu) [process, left of=check_mu, xshift=-5cm] {Set $\mu_{\rm outflow}\leftarrow\mu_{\rm outflow,new}$ and repeat};

    \node (end) [startstop, below of=check_mu, yshift=-1cm] {Output self--consistent $(\dot M,R_{\rm XUV},c_s)$ and $(\phi_{\rm H},\phi_{\rm O},x_{\rm O})$};

    \draw [arrow] (start) -- (set_mu);
    \draw [arrow] (set_mu) -- (massloss_init);
    \draw [arrow] (massloss_init) -- (compute_el);
    \draw [arrow] (compute_el) -- (guess_cs);
    \draw [arrow] (guess_cs) -- (check_cs);
    \draw [arrow] (check_cs.east) -- node[midway, above] {No} (adjust_cs.west);
    \draw [arrow] (adjust_cs) |- (guess_cs.east);
    \draw [arrow] (check_cs.south) -- node[midway, left] {Yes} (update_reuv);
    \draw [arrow] (update_reuv) -- (massloss_done);

    \draw [arrow] (massloss_done) -- (classify);
    \draw [arrow] (classify.south) -- node[midway, above] {No} (el_branch);
    \draw [arrow] (classify.south) -- node[midway, above] {Yes} (rl_branch);
    \draw [arrow] (el_branch) -- (select_branch);
    \draw [arrow] (rl_branch) -- (select_branch);

    \draw [arrow] (select_branch) -- (calc_tout);
    \draw [arrow] (calc_tout) -- (calc_fmass);
    \draw [arrow] (calc_fmass) -- (frac_solve);
    \draw [arrow] (frac_solve) -- (update_mu);
    \draw [arrow] (update_mu) -- (check_mu);

    \draw [arrow] (check_mu.west) -- node[midway, above] {No} (iter_mu.east);
    \draw [arrow] (iter_mu) |- (set_mu.west);

    \draw [arrow] (check_mu.south) -- node[midway, left] {Yes} (end);

    \end{tikzpicture}
\end{figure}
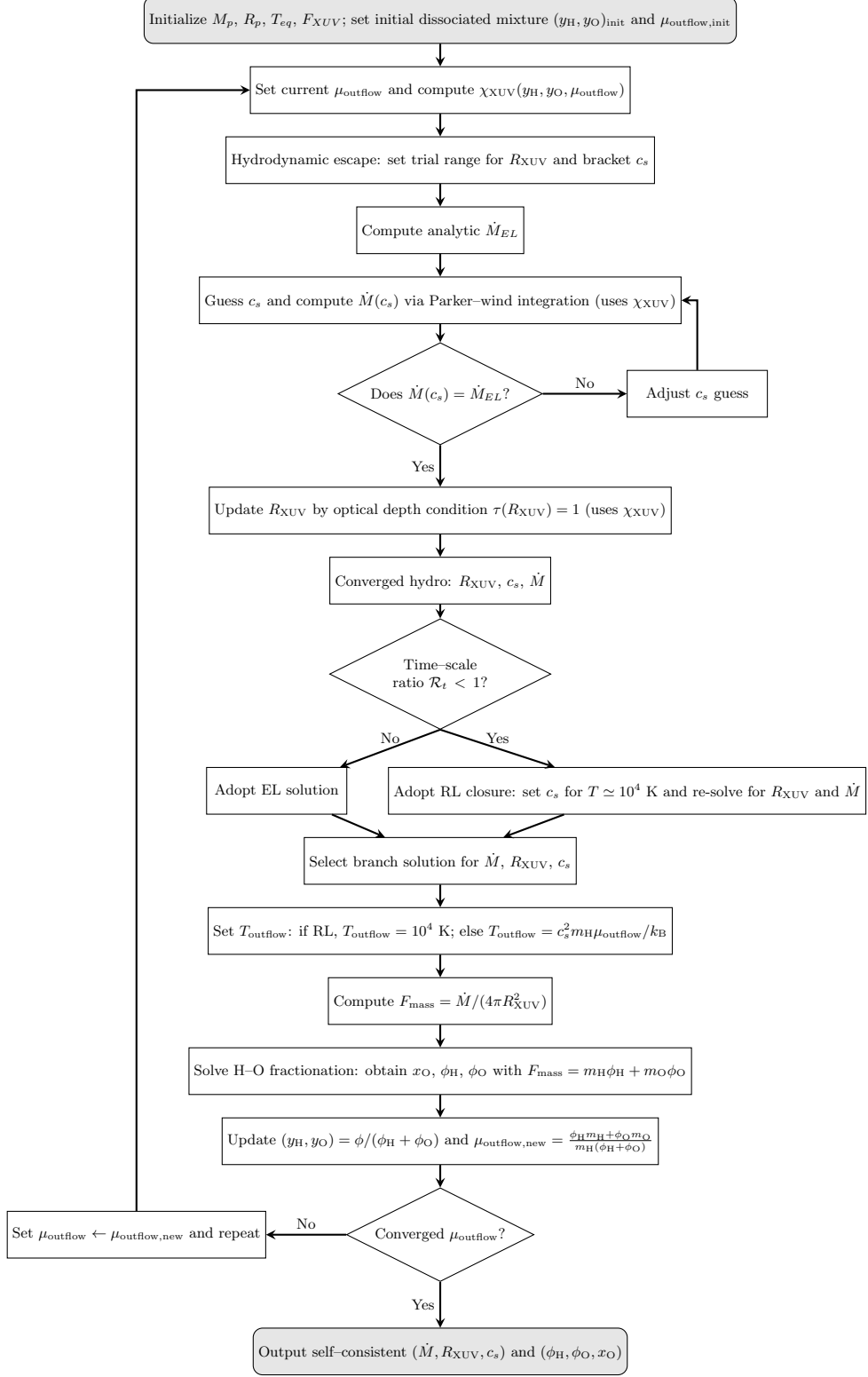


\bibliography{Bibliography}{}

@article{zahnle_mass_1986,
	title = {Mass fractionation during transonic escape and implications for loss of water from {Mars} and {Venus}},
	volume = {68},
	issn = {0019-1035},
	url = {https://www.sciencedirect.com/science/article/pii/0019103586900515},
	doi = {10.1016/0019-1035(86)90051-5},
	number = {3},
	journal = {Icarus},
	author = {Zahnle, Kevin J. and Kasting, James F.},
	month = dec,
	year = {1986},
	pages = {462--480},
}

@article{steinmeyer2026coupled,
  title={Coupled thermal-chemical evolution models of sub-Neptunes reveal atmospheric signatures of their formation location},
  author={Steinmeyer, Marie-Luise and Dorn, Caroline and Werlen, Aaron and Grimm, Simon L},
  journal={arXiv preprint arXiv:2601.21377},
  year={2026}
}

@article{rogers2026using,
  title={Using observations of escaping H/He to constrain the atmospheric composition of sub-Neptunes},
  author={Rogers, James G and Owen, James E and Schreyer, Ethan and Kirk, James},
  journal={arXiv preprint arXiv:2601.14254},
  year={2026}
}

@article{dorn2015can,
  title={Can we constrain the interior structure of rocky exoplanets from mass and radius measurements?},
  author={Dorn, Caroline and Khan, Amir and Heng, Kevin and Connolly, James AD and Alibert, Yann and Benz, Willy and Tackley, Paul},
  journal={Astronomy \& Astrophysics},
  volume={577},
  pages={A83},
  year={2015},
  publisher={EDP Sciences}
}

@article{dorn2021hidden,
  title={Hidden water in magma ocean exoplanets},
  author={Dorn, Caroline and Lichtenberg, Tim},
  journal={The Astrophysical Journal Letters},
  volume={922},
  number={1},
  pages={L4},
  year={2021},
  publisher={IOP Publishing}
}

@article{Kite_2021,
doi = {10.3847/2041-8213/abe7dc},
url = {https://doi.org/10.3847/2041-8213/abe7dc},
year = {2021},
month = {mar},
publisher = {The American Astronomical Society},
volume = {909},
number = {2},
pages = {L22},
author = {Kite, Edwin S. and Schaefer, Laura},
title = {Water on Hot Rocky Exoplanets},
journal = {The Astrophysical Journal Letters},
}

@article{Seo_2024,
doi = {10.3847/1538-4357/ad7461},
url = {https://doi.org/10.3847/1538-4357/ad7461},
year = {2024},
month = {oct},
publisher = {The American Astronomical Society},
volume = {975},
number = {1},
pages = {14},
author = {Seo, Chanoul and Ito, Yuichi and Fujii, Yuka},
title = {Role of Magma Oceans in Controlling Carbon and Oxygen of Sub-Neptune Atmospheres},
journal = {The Astrophysical Journal},
}

@article{owen_mapping_2023,
       author = {{Owen}, James E. and {Schlichting}, Hilke E.},
        title = "{Mapping out the parameter space for photoevaporation and core-powered mass-loss}",
      journal = {\mnras},
         year = 2024,
        month = feb,
       volume = {528},
       number = {2},
        pages = {1615-1629},
          doi = {10.1093/mnras/stad3972},
archivePrefix = {arXiv},
       eprint = {2308.00020},
 primaryClass = {astro-ph.EP},
       adsurl = {https://ui.adsabs.harvard.edu/abs/2024MNRAS.528.1615O}
}

@misc{luo_majority_2024,
	title = {Majority of water hides deep in the interiors of exoplanets},
	url = {http://arxiv.org/abs/2401.16394},
	doi = {10.48550/arXiv.2401.16394},
	publisher = {arXiv},
	author = {Luo, Haiyang and Dorn, Caroline and Deng, Jie},
	month = jan,
	year = {2024},
	note = {arXiv:2401.16394 [astro-ph]},
}

@article{murray_atmospheric_2009,
  title = {ATMOSPHERIC ESCAPE FROM HOT JUPITERS},
  volume = {693},
  ISSN = {1538-4357},
  url = {http://dx.doi.org/10.1088/0004-637X/693/1/23},
  DOI = {10.1088/0004-637x/693/1/23},
  number = {1},
  journal = {The Astrophysical Journal},
  publisher = {American Astronomical Society},
  author = {Murray-Clay,  Ruth A. and Chiang,  Eugene I. and Murray,  Norman},
  year = {2009},
  month = feb,
  pages = {23–42}
}

@article{kawamura_reduced_2024,
	title = {Reduced {Water} {Loss} due to {Photochemistry} on {Terrestrial} {Planets} in the {Runaway} {Greenhouse} {Phase} around {Pre}-main-sequence {M} {Dwarfs}},
	volume = {967},
	issn = {0004-637X, 1538-4357},
	url = {https://iopscience.iop.org/article/10.3847/1538-4357/ad3e7e},
	doi = {10.3847/1538-4357/ad3e7e},
	language = {en},
	number = {2},
	urldate = {2024-06-10},
	journal = {The Astrophysical Journal},
	author = {Kawamura, Yo and Yoshida, Tatsuya and Terada, Naoki and Nakamura, Yuki and Koyama, Shungo and Karyu, Hiroki and Terada, Kaori and Sakai, Shotaro},
	month = jun,
	year = {2024},
	pages = {95},
}

@misc{broussard_impact_2025,
	title = {The {Impact} of {Extended} {CO}\$\_2\$ {Cross} {Sections} on {Temperate} {Anoxic} {Planet} {Atmospheres}},
	url = {http://arxiv.org/abs/2501.08434},
	doi = {10.48550/arXiv.2501.08434},
	urldate = {2025-01-16},
	publisher = {arXiv},
	author = {Broussard, Wynter and Schwieterman, Edward W. and Sousa-Silva, Clara and Sanger-Johnson, Grace and Ranjan, Sukrit and Venot, Olivia},
	month = jan,
	year = {2025},
	note = {arXiv:2501.08434 [astro-ph]},
}

@misc{yoshida_suppression_2024,
	title = {Suppression of hydrodynamic escape of an {H2}-rich early {Earth} atmosphere by radiative cooling of carbon oxides},
	url = {http://arxiv.org/abs/2411.15456},
	doi = {10.48550/arXiv.2411.15456},
	urldate = {2024-11-27},
	publisher = {arXiv},
	author = {Yoshida, Tatsuya and Terada, Naoki and Kuramoto, Kiyoshi},
	month = nov,
	year = {2024},
	note = {arXiv:2411.15456},
}

@article{zahnle_mass_1990,
	title = {Mass fractionation of noble gases in diffusion-limited hydrodynamic hydrogen escape},
	volume = {84},
	issn = {0019-1035},
	url = {https://www.sciencedirect.com/science/article/pii/001910359090050J},
	doi = {10.1016/0019-1035(90)90050-J},
	number = {2},
	journal = {Icarus},
	author = {Zahnle, Kevin and Kasting, James F. and Pollack, James B.},
	month = apr,
	year = {1990},
	pages = {502--527},
}

@article{hunten_mass_1987,
	title = {Mass fractionation in hydrodynamic escape},
	volume = {69},
	issn = {0019-1035},
	url = {https://www.sciencedirect.com/science/article/pii/0019103587900224},
	doi = {10.1016/0019-1035(87)90022-4},
	number = {3},
	journal = {Icarus},
	author = {Hunten, Donald M. and Pepin, Robert O. and Walker, James C. G.},
	month = mar,
	year = {1987},
	pages = {532--549},
}

@article{zahnle_photochemistry_1986,
	title = {Photochemistry of methane and the formation of hydrocyanic acid ({HCN}) in the {Earth}'s early atmosphere},
	volume = {91},
	copyright = {Copyright 1986 by the American Geophysical Union.},
	issn = {2156-2202},
	doi = {10.1029/JD091iD02p02819},
	language = {en},
	number = {D2},
	urldate = {2025-11-13},
	journal = {Journal of Geophysical Research: Atmospheres},
	author = {Zahnle, Kevin J.},
	year = {1986},
	note = {\_eprint: https://agupubs.onlinelibrary.wiley.com/doi/pdf/10.1029/JD091iD02p02819},
	pages = {2819--2834},
}

@misc{cherubim_strong_2024,
	title = {Strong fractionation of deuterium and helium in sub-{Neptune} atmospheres along the radius valley},
	url = {http://arxiv.org/abs/2402.10690},
	doi = {10.48550/arXiv.2402.10690},
	urldate = {2024-02-27},
	publisher = {arXiv},
	author = {Cherubim, Collin and Wordsworth, Robin and Hu, Renyu and Shkolnik, Evgenya},
	month = feb,
	year = {2024},
	note = {arXiv:2402.10690 [astro-ph]},
}

@article{loyd_current_2020,
	title = {Current {Population} {Statistics} {Do} {Not} {Favor} {Photoevaporation} over {Core}-powered {Mass} {Loss} as the {Dominant} {Cause} of the {Exoplanet} {Radius} {Gap}},
	volume = {890},
	issn = {0004-637X},
	url = {https://dx.doi.org/10.3847/1538-4357/ab6605},
	doi = {10.3847/1538-4357/ab6605},
	language = {en},
	number = {1},
	urldate = {2024-04-22},
	journal = {The Astrophysical Journal},
	author = {Loyd, R. O. Parke and Shkolnik, Evgenya L. and Schneider, Adam C. and Richey-Yowell, Tyler and Barman, Travis S. and Peacock, Sarah and Pagano, Isabella},
	month = feb,
	year = {2020},
	note = {Publisher: The American Astronomical Society},
	pages = {23},
}

@article{lopez_how_2012,
	title = {How {Thermal} {Evolution} and {Mass}-loss {Sculpt} {Populations} of {Super}-{Earths} and {Sub}-{Neptunes}: {Application} to the {Kepler}-11 {System} and {Beyond}},
	volume = {761},
	issn = {0004-637X},
	shorttitle = {How {Thermal} {Evolution} and {Mass}-loss {Sculpt} {Populations} of {Super}-{Earths} and {Sub}-{Neptunes}},
	url = {https://ui.adsabs.harvard.edu/abs/2012ApJ...761...59L},
	doi = {10.1088/0004-637X/761/1/59},
	urldate = {2025-02-18},
	journal = {The Astrophysical Journal},
	author = {Lopez, Eric D. and Fortney, Jonathan J. and Miller, Neil},
	month = dec,
	year = {2012},
	note = {Publisher: IOP
ADS Bibcode: 2012ApJ...761...59L},
	pages = {59},
	}

@article{jin_planetary_2014,
	title = {Planetary population synthesis coupled with atmospheric escape: a statistical view of evaporation},
	volume = {795},
	issn = {1538-4357},
	shorttitle = {Planetary population synthesis coupled with atmospheric escape},
	url = {http://arxiv.org/abs/1409.2879},
	doi = {10.1088/0004-637X/795/1/65},
	number = {1},
	urldate = {2025-02-18},
	journal = {The Astrophysical Journal},
	author = {Jin, Sheng and Mordasini, Christoph and Parmentier, Vivien and Boekel, Roy van and Henning, Thomas and Ji, Jianghui},
	month = oct,
	year = {2014},
	note = {arXiv:1409.2879 [astro-ph]},
	pages = {65},
}

@article{rogers_photoevaporation_2021,
	title = {Photoevaporation vs. core-powered mass-loss: model comparison with the {3D} radius gap},
	volume = {508},
	issn = {0035-8711, 1365-2966},
	shorttitle = {Photoevaporation vs. core-powered mass-loss},
	url = {http://arxiv.org/abs/2105.03443},
	doi = {10.1093/mnras/stab2897},
	number = {4},
	urldate = {2024-02-27},
	journal = {Monthly Notices of the Royal Astronomical Society},
	author = {Rogers, James G. and Gupta, Akash and Owen, James E. and Schlicdornhting, Hilke E.},
	month = nov,
	year = {2021},
	note = {arXiv:2105.03443 [astro-ph]},
	pages = {5886--5902},
}

@article{jiang_estimating_2025,
	title = {Estimating the {Mass} {Escaping} {Rates} of {Radius}-valley-spanning {Planets} in the {TOI}-431 {System} via {X}-{Ray} and {Ultraviolet} {Evaporation}},
	volume = {980},
	issn = {0004-637X, 1538-4357},
	url = {http://arxiv.org/abs/2502.07294},
	doi = {10.3847/1538-4357/ada607},
	number = {2},
	urldate = {2025-02-19},
	journal = {The Astrophysical Journal},
	author = {Jiang, Xiaoming and Jiang, Jonathan H. and Burn, Remo and Zhu, Zong-Hong},
	month = feb,
	year = {2025},
	note = {arXiv:2502.07294 [astro-ph]},
	pages = {175},
}

@article{baraffe_new_2015,
	title = {New evolutionary models for pre-main sequence and main sequence low-mass stars down to the hydrogen-burning limit},
	volume = {577},
	issn = {0004-6361, 1432-0746},
	url = {http://arxiv.org/abs/1503.04107},
	doi = {10.1051/0004-6361/201425481},
	urldate = {2025-05-05},
	journal = {Astronomy \& Astrophysics},
	author = {Baraffe, I. and Homeier, D. and Allard, F. and Chabrier, G.},
	month = may,
	year = {2015},
	note = {arXiv:1503.04107 [astro-ph]},
	pages = {A42},
}

@misc{burn_water-rich_2024,
	title = {Water-rich sub-{Neptunes} and rocky super {Earths} around different {Stars}: {Radii} shaped by {Volatile} {Partitioning}, {Formation}, and {Evolution}},
	shorttitle = {Water-rich sub-{Neptunes} and rocky super {Earths} around different {Stars}},
	url = {http://arxiv.org/abs/2411.16879},
	doi = {10.48550/arXiv.2411.16879},
	urldate = {2025-05-19},
	publisher = {arXiv},
	author = {Burn, Remo and Bali, Komal and Dorn, Caroline and Luque, Rafael and Grimm, Simon L.},
	month = nov,
	year = {2024},
	note = {arXiv:2411.16879 [astro-ph]},
}

@article{luger_extreme_2015,
	title = {Extreme {Water} {Loss} and {Abiotic} {O}\$\_2\$ {Buildup} {On} {Planets} {Throughout} the {Habitable} {Zones} of {M} {Dwarfs}},
	volume = {15},
	issn = {1531-1074, 1557-8070},
	url = {http://arxiv.org/abs/1411.7412},
	doi = {10.1089/ast.2014.1231},
	number = {2},
	urldate = {2025-05-19},
	journal = {Astrobiology},
	author = {Luger, Rodrigo and Barnes, Rory},
	month = feb,
	year = {2015},
	note = {arXiv:1411.7412 [astro-ph]},
	pages = {119--143},
}

@article{dorn_generalized_2017,
	title = {A generalized bayesian inference method for constraining the interiors of super {Earths} and sub-{Neptunes}},
	volume = {597},
	issn = {0004-6361, 1432-0746},
	url = {http://arxiv.org/abs/1609.03908},
	doi = {10.1051/0004-6361/201628708},
	urldate = {2025-05-19},
	journal = {Astronomy \& Astrophysics},
	author = {Dorn, C. and Venturini, J. and Khan, A. and Heng, K. and Alibert, Y. and Helled, R. and Rivoldini, A. and Benz, W.},
	month = jan, 
	year = {2017},
	note = {arXiv:1609.03908 [astro-ph]},
	pages = {A37},
}

@article{osborn_toi-431hip_2021,
	title = {{TOI}-431/{HIP} 26013: a super-{Earth} and a sub-{Neptune} transiting a bright, early {K} dwarf, with a third {RV} planet},
	volume = {507},
	issn = {0035-8711, 1365-2966},
	shorttitle = {{TOI}-431/{HIP} 26013},
	url = {http://arxiv.org/abs/2108.02310},
	doi = {10.1093/mnras/stab2313},
	number = {2},
	urldate = {2025-05-21},
	journal = {Monthly Notices of the Royal Astronomical Society},
	author = {Osborn, Ares and Armstrong, David J. and Cale, Bryson and Brahm, Rafael and Wittenmyer, Robert A. and Dai, Fei and Crossfield, Ian J. M. and Bryant, Edward M. and Adibekyan, Vardan and Cloutier, Ryan and Collins, Karen A. and Mena, E. Delgado and Fridlund, Malcolm and Hellier, Coel and Howell, Steve B. and King, George W. and Lillo-Box, Jorge and Otegi, Jon and Sousa, S. and Stassun, Keivan G. and Matthews, Elisabeth C. and Ziegler, Carl and Ricker, George and Vanderspek, Roland and Latham, David W. and Seager, S. and Winn, Joshua N. and Jenkins, Jon M. and Acton, Jack S. and Addison, Brett C. and Anderson, David R. and Ballard, Sarah and Barrado, David and Barros, Susana C. C. and Batalha, Natalie and Bayliss, Daniel and Barclay, Thomas and Benneke, Björn and Jr, John Berberian and Bouchy, Francois and Bowler, Brendan P. and Briceño, César and Burke, Christopher J. and Burleigh, Matthew R. and Casewell, Sarah L. and Ciardi, David and Collins, Kevin I. and Cooke, Benjamin F. and Demangeon, Olivier D. S. and Díaz, Rodrigo F. and Dorn, C. and Dragomir, Diana and Dressing, Courtney and Dumusque, Xavier and Espinoza, Néstor and Figueira, P. and Fulton, Benjamin and Furlan, E. and Gaidos, E. and Geneser, C. and Gill, Samuel and Goad, Michael R. and Gonzales, Erica J. and Gorjian, Varoujan and Günther, Maximilian N. and Helled, Ravit and Henderson, Beth A. and Henning, Thomas and Hogan, Aleisha and Hojjatpanah, Saeed and Horner, Jonathan and Howard, Andrew W. and Hoyer, Sergio and Huber, Dan and Isaacson, Howard and Jenkins, James S. and Jensen, Eric L. N. and Jordán, Andrés and Kane, Stephen R. and Jr, Richard C. Kidwell and Kielkopf, John and Law, Nicholas and Lendl, Monika and Lund, M. and Matson, Rachel A. and Mann, Andrew W. and McCormac, James and Mengel, Matthew W. and Morales, Farisa Y. and Nielsen, Louise D. and Okumura, Jack and Osborn, Hugh P. and Petigura, Erik A. and Plavchan, Peter and Pollacco, Don and Quintana, Elisa V. and Raynard, Liam and Robertson, Paul and Rose, Mark E. and Roy, Arpita and Reefe, Michael and Santerne, Alexandre and Santos, Nuno C. and Sarkis, Paula and Schlieder, J. and Schwarz, Richard P. and Scott, Nicholas J. and Shporer, Avi and Smith, A. M. S. and Stibbard, C. and Stockdale, Chris and Strøm, Paul A. and Twicken, Joseph D. and Tan, Thiam-Guan and Tanner, A. and Teske, J. and Tilbrook, Rosanna H. and Tinney, C. G. and Udry, Stephane and Villaseñor, Jesus Noel and Vines, Jose I. and Wang, Sharon X. and Weiss, Lauren M. and West, Richard G. and Wheatley, Peter J. and Wright, Duncan J. and Zhang, Hui and Zohrabi, F.},
	month = sep,
	year = {2021},
	note = {arXiv:2108.02310 [astro-ph]},
	pages = {2782--2803},
}

@article{piaulet-ghorayeb_jwstniriss_2024,
	title = {{JWST}/{NIRISS} reveals the water-rich "steam world" atmosphere of {GJ} 9827 d},
	volume = {974},
	issn = {2041-8205, 2041-8213},
	url = {http://arxiv.org/abs/2410.03527},
	doi = {10.3847/2041-8213/ad6f00},
	number = {1},
	urldate = {2025-06-04},
	journal = {The Astrophysical Journal Letters},
	author = {Piaulet-Ghorayeb, Caroline and Benneke, Bjorn and Radica, Michael and Raul, Eshan and Coulombe, Louis-Philippe and Ahrer, Eva-Maria and Kubyshkina, Daria and Howard, Ward S. and Krissansen-Totton, Joshua and MacDonald, Ryan and Roy, Pierre-Alexis and Louca, Amy and Christie, Duncan and Fournier-Tondreau, Marylou and Allart, Romain and Miguel, Yamila and Schlichting, Hilke E. and Welbanks, Luis and Cadieux, Charles and Dorn, Caroline and Evans-Soma, Thomas M. and Fortney, Jonathan J. and Pierrehumbert, Raymond and Lafreniere, David and Acuna, Lorena and Komacek, Thaddeus and Innes, Hamish and Beatty, Thomas G. and Cloutier, Ryan and Doyon, Rene and Gagnebin, Anna and Gapp, Cyril and Knutson, Heather A.},
	month = oct,
	year = {2024},
	note = {arXiv:2410.03527 [astro-ph]},
	pages = {L10},
}

@article{bean_nature_2021,
	title = {The nature and origins of sub-{Neptune} size planets},
	volume = {126},
	issn = {2169-9097, 2169-9100},
	url = {http://arxiv.org/abs/2010.11867},
	doi = {10.1029/2020JE006639},
	number = {1},
	urldate = {2025-06-04},
	journal = {Journal of Geophysical Research: Planets},
	author = {Bean, Jacob L. and Raymond, Sean N. and Owen, James E.},
	month = jan,
	year = {2021},
	note = {arXiv:2010.11867 [astro-ph]},
	pages = {e2020JE006639},
}

@incollection{lammer_atmospheric_2008,
	address = {New York, NY},
	title = {Atmospheric {Escape} and {Evolution} of {Terrestrial} {Planets} and {Satellites}},
	isbn = {978-0-387-87825-6},
	url = {https://doi.org/10.1007/978-0-387-87825-6_11},
	language = {en},
	urldate = {2025-06-10},
	booktitle = {Comparative {Aeronomy}},
	publisher = {Springer},
	author = {Lammer, Helmut and Kasting, James F. and Chassefière, Eric and Johnson, Robert E. and Kulikov, Yuri N. and Tian, Feng},
	editor = {Nagy, Andrew F. and Balogh, André and Cravens, Thomas E. and Mendillo, Michael and Mueller-Wodarg, Ingo},
	year = {2008},
	doi = {10.1007/978-0-387-87825-6_11},
	pages = {399--436},
}

@article{owen_atmospheric_2019,
	title = {Atmospheric {Escape} and the {Evolution} of {Close}-{In} {Exoplanets}},
	volume = {47},
	issn = {0084-6597, 1545-4495},
	url = {https://www.annualreviews.org/content/journals/10.1146/annurev-earth-053018-060246},
	doi = {10.1146/annurev-earth-053018-060246},
	language = {en},
	number = {Volume 47, 2019},
	urldate = {2025-06-10},
	journal = {Annual Review of Earth and Planetary Sciences},
	author = {Owen, James E.},
	month = may,
	year = {2019},
	note = {Publisher: Annual Reviews},
	pages = {67--90},
}

@article{tian_atmospheric_2015,
	title = {Atmospheric {Escape} from {Solar} {System} {Terrestrial} {Planets} and {Exoplanets}},
	volume = {43},
	issn = {0084-6597},
	url = {https://ui.adsabs.harvard.edu/abs/2015AREPS..43..459T},
	doi = {10.1146/annurev-earth-060313-054834},
	urldate = {2025-06-10},
	journal = {Annual Review of Earth and Planetary Sciences},
	author = {Tian, Feng},
	month = may,
	year = {2015},
	note = {ADS Bibcode: 2015AREPS..43..459T},
	pages = {459--476},
}

@article{eylen_masses_2021,
	title = {Masses and compositions of three small planets orbiting the nearby {M} dwarf {L231}-32 ({TOI}-270) and the {M} dwarf radius valley},
	issn = {0035-8711, 1365-2966},
	url = {http://arxiv.org/abs/2101.01593},
	doi = {10.1093/mnras/stab2143},
	urldate = {2025-06-10},
	journal = {Monthly Notices of the Royal Astronomical Society},
	author = {Eylen, Vincent Van and Astudillo-Defru, N. and Bonfils, X. and Livingston, J. and Hirano, T. and Luque, R. and Lam, K. W. F. and Justesen, A. B. and Winn, J. N. and Gandolfi, D. and Nowak, G. and Palle, E. and Albrecht, S. and Dai, F. and Estrada, B. Campos and Owen, J. E. and Foreman-Mackey, D. and Fridlund, M. and Korth, J. and Mathur, S. and Forveille, T. and Mikal-Evans, T. and Osborne, H. L. M. and Ho, C. S. K. and Almenara, J. M. and Artigau, E. and Barragán, O. and Barros, S. C. C. and Bouchy, F. and Cabrera, J. and Caldwell, D. A. and Charbonneau, D. and Chaturvedi, P. and Cochran, W. D. and Csizmadia, S. and Damasso, M. and Delfosse, X. and Medeiros, J. R. De and Díaz, R. F. and Doyon, R. and Esposito, M. and Fűrész, G. and Figueira, P. and Georgieva, I. and Goffo, E. and Grziwa, S. and Guenther, E. and Hatzes, A. P. and Jenkins, J. M. and Kabath, P. and Knudstrup, E. and Latham, D. W. and Lavie, B. and Lovis, C. and Mennickent, R. E. and Mullally, S. E. and Murgas, F. and Narita, N. and Pepe, F. A. and Persson, C. M. and Redfield, S. and Ricker, G. R. and Santos, N. C. and Seager, S. and Serrano, L. M. and Smith, A. M. S. and Mascareño, A. Suárez and Subjak, J. and Twicken, J. D. and Udry, S. and Vanderspek, R. and Osorio, M. R. Zapatero},
	month = aug,
	year = {2021},
	note = {arXiv:2101.01593 [astro-ph]},
	pages = {stab2143},
}

@article{madhusudhan_carbon-bearing_2023,
	title = {Carbon-bearing {Molecules} in a {Possible} {Hycean} {Atmosphere}},
	volume = {956},
	issn = {2041-8205, 2041-8213},
	url = {https://iopscience.iop.org/article/10.3847/2041-8213/acf577},
	doi = {10.3847/2041-8213/acf577},
	language = {en},
	number = {1},
	urldate = {2024-06-05},
	journal = {The Astrophysical Journal Letters},
	author = {Madhusudhan, Nikku and Sarkar, Subhajit and Constantinou, Savvas and Holmberg, Måns and Piette, Anjali A. A. and Moses, Julianne I.},
	month = oct,
	year = {2023},
	pages = {L13},
}

@article{owen_uv_2015,
	title = {{UV} {DRIVEN} {EVAPORATION} {OF} {CLOSE}-{IN} {PLANETS}: {ENERGY}-{LIMITED}, {RECOMBINATION}-{LIMITED}, {AND} {PHOTON}-{LIMITED} {FLOWS}},
	volume = {816},
	issn = {0004-637X},
	shorttitle = {{UV} {DRIVEN} {EVAPORATION} {OF} {CLOSE}-{IN} {PLANETS}},
	url = {https://dx.doi.org/10.3847/0004-637X/816/1/34},
	doi = {10.3847/0004-637X/816/1/34},
	language = {en},
	number = {1},
	urldate = {2025-06-10},
	journal = {The Astrophysical Journal},
	author = {Owen, James E. and Alvarez, Marcelo A.},
	month = dec,
	year = {2015},
	note = {Publisher: The American Astronomical Society},
	pages = {34},
}

@article{ginzburg_core-powered_2018,
	title = {Core-powered mass-loss and the radius distribution of small exoplanets},
	volume = {476},
	issn = {0035-8711},
	url = {https://doi.org/10.1093/mnras/sty290},
	doi = {10.1093/mnras/sty290},
	number = {1},
	urldate = {2025-06-10},
	journal = {Monthly Notices of the Royal Astronomical Society},
	author = {Ginzburg, Sivan and Schlichting, Hilke E and Sari, Re'em},
	month = may,
	year = {2018},
	pages = {759--765},
}

@article{lammer_atmospheric_2003,
	title = {Atmospheric {Loss} of {Exoplanets} {Resulting} from {Stellar} {X}-{Ray} and {Extreme}-{Ultraviolet} {Heating}},
	volume = {598},
	issn = {0004-637X},
	url = {https://iopscience.iop.org/article/10.1086/380815/meta},
	doi = {10.1086/380815},
	language = {en},
	number = {2},
	urldate = {2025-06-10},
	journal = {The Astrophysical Journal},
	author = {Lammer, H. and Selsis, F. and Ribas, I. and Guinan, E. F. and Bauer, S. J. and Weiss, W. W.},
	month = nov,
	year = {2003},
	note = {Publisher: IOP Publishing},
	pages = {L121},
}

@misc{ballabio_understanding_2025,
	title = {Understanding what helium absorption tells us about atmospheric escape from exoplanets},
	url = {http://arxiv.org/abs/2501.06149},
	doi = {10.48550/arXiv.2501.06149},
	publisher = {arXiv},
	author = {Ballabio, Giulia and Owen, James E.},
	month = jan,
	year = {2025},
	note = {arXiv:2501.06149 [astro-ph]},
}

@article{becker_coupled_2020,
	title = {A {Coupled} {Analysis} of {Atmospheric} {Mass} {Loss} and {Tidal} {Evolution} in {XUV} {Irradiated} {Exoplanets}: {The} {TRAPPIST}-1 {Case} {Study}},
	volume = {159},
	issn = {1538-3881},
	shorttitle = {A {Coupled} {Analysis} of {Atmospheric} {Mass} {Loss} and {Tidal} {Evolution} in {XUV} {Irradiated} {Exoplanets}},
	url = {https://dx.doi.org/10.3847/1538-3881/ab8fb0},
	doi = {10.3847/1538-3881/ab8fb0},
	language = {en},
	number = {6},
	journal = {The Astronomical Journal},
	author = {Becker, Juliette and Gallo, Elena and Hodges-Kluck, Edmund and Adams, Fred C. and Barnes, Rory},
	month = may,
	year = {2020},
	note = {Publisher: The American Astronomical Society},
	pages = {275},
}

@article{santos_high-energy_2020,
	title = {The high-energy environment and atmospheric escape of the mini-{Neptune} {K2}-18 b},
	volume = {634},
	issn = {0004-6361, 1432-0746},
	url = {http://arxiv.org/abs/2001.04532},
	doi = {10.1051/0004-6361/201937327},
	urldate = {2025-06-18},
	journal = {Astronomy \& Astrophysics},
	author = {Santos, Leonardo A. dos and Ehrenreich, David and Bourrier, Vincent and Astudillo-Defru, Nicola and Bonfils, Xavier and Forget, François and Lovis, Christophe and Pepe, Francesco and Udry, Stéphane},
	month = feb,
	year = {2020},
	note = {arXiv:2001.04532 [astro-ph]},
	pages = {L4},
}

@article{zhang_detection_2022,
	title = {Detection of {Ongoing} {Mass} {Loss} from {HD} 63433c, a {Young} {Mini}-{Neptune}},
	volume = {163},
	issn = {1538-3881},
	url = {https://dx.doi.org/10.3847/1538-3881/ac3f3b},
	doi = {10.3847/1538-3881/ac3f3b},
	language = {en},
	number = {2},
	urldate = {2025-06-18},
	journal = {The Astronomical Journal},
	author = {Zhang, Michael and Knutson, Heather A. and Wang, Lile and Dai, Fei and dos Santos, Leonardo A. and Fossati, Luca and Henry, Gregory W. and Ehrenreich, David and Alibert, Yann and Hoyer, Sergio and Wilson, Thomas G. and Bonfanti, Andrea},
	month = jan,
	year = {2022},
	note = {Publisher: The American Astronomical Society},
	pages = {68},
}

@article{Zhang2023-kc,
  title     = "Detection of atmospheric escape from four young mini-Neptunes",
  author    = "Zhang, Michael and Knutson, Heather A and Dai, Fei and Wang,
               Lile and Ricker, George R and Schwarz, Richard P and Mann,
               Christopher and Collins, Karen",
  journal   = "Astron. J.",
  publisher = "American Astronomical Society",
  volume    =  165,
  number    =  2,
  pages     = "62",
  month     =  feb,
  year      =  2023,
  copyright = "http://creativecommons.org/licenses/by/4.0/"
}

@misc{werlen_sub-neptunes_2025,
	title = {Sub-{Neptunes} {Are} {Drier} {Than} {They} {Seem}: {Rethinking} the {Origins} of {Water}-{Rich} {Worlds}},
	shorttitle = {Sub-{Neptunes} {Are} {Drier} {Than} {They} {Seem}},
	url = {http://arxiv.org/abs/2507.00765},
	doi = {10.48550/arXiv.2507.00765},
	urldate = {2025-07-02},
	publisher = {arXiv},
	author = {Werlen, Aaron and Dorn, Caroline and Burn, Remo and Schlichting, Hilke E. and Grimm, Simon L. and Young, Edward D.},
	month = jul,
	year = {2025},
	note = {arXiv:2507.00765 [astro-ph]},
}

@misc{rogers_most_2024,
	title = {Most {Super}-{Earths} {Have} {Less} {Than} 3\% {Water}},
	url = {http://arxiv.org/abs/2409.17394},
	doi = {10.48550/arXiv.2409.17394},
	urldate = {2025-07-09},
	publisher = {arXiv},
	author = {Rogers, James G. and Dorn, Caroline and Raj, Vivasvaan Aditya and Schlichting, Hilke E. and Young, Edward D.},
	month = dec,
	year = {2024},
	note = {arXiv:2409.17394 [astro-ph]},
}

@misc{mordasini_planetary_2020,
	title = {Planetary evolution with atmospheric photoevaporation {I}. {Analytical} derivation and numerical study of the evaporation valley and transition from super-{Earths} to sub-{Neptunes}},
	url = {http://arxiv.org/abs/2002.02455},
	doi = {10.1051/0004-6361/201935541},
	urldate = {2025-07-09},
	author = {Mordasini, Christoph},
	month = feb,
	year = {2020},
	note = {arXiv:2002.02455 [astro-ph]},
}

@article{owen_evaporation_2017,
	title = {The {Evaporation} {Valley} in the {Kepler} {Planets}},
	volume = {847},
	issn = {0004-637X},
	url = {https://dx.doi.org/10.3847/1538-4357/aa890a},
	doi = {10.3847/1538-4357/aa890a},
	language = {en},
	number = {1},
	urldate = {2025-07-10},
	journal = {The Astrophysical Journal},
	author = {Owen, James E. and Wu, Yanqin},
	month = sep,
	year = {2017},
	note = {Publisher: The American Astronomical Society},
	pages = {29},
}

@article{owen_atmospheres_2016,
	title = {Atmospheres of low-mass planets: the "boil-off"},
	volume = {817},
	issn = {0004-637X, 1538-4357},
	shorttitle = {Atmospheres of low-mass planets},
	url = {http://arxiv.org/abs/1506.02049},
	doi = {10.3847/0004-637X/817/2/107},
	number = {2},
	urldate = {2025-07-10},
	journal = {The Astrophysical Journal},
	author = {Owen, James E. and Wu, Yanqin},
	month = feb,
	year = {2016},
	note = {arXiv:1506.02049 [astro-ph]},
	pages = {107},
}

@misc{guillot_radiative_2010,
	title = {On the radiative equilibrium of irradiated planetary atmospheres},
	url = {http://arxiv.org/abs/1006.4702},
	doi = {10.1051/0004-6361/200913396},
	urldate = {2025-07-10},
	author = {Guillot, Tristan},
	month = jun,
	year = {2010},
	note = {arXiv:1006.4702 [astro-ph]},
}

@article{owen_planetary_2012,
	title = {Planetary evaporation by {UV} \& {X}-ray radiation: basic hydrodynamics},
	volume = {425},
	issn = {00358711},
	shorttitle = {Planetary evaporation by {UV} \& {X}-ray radiation},
	url = {http://arxiv.org/abs/1206.2367},
	doi = {10.1111/j.1365-2966.2012.21481.x},
	number = {4},
	urldate = {2024-04-25},
	journal = {Monthly Notices of the Royal Astronomical Society},
	author = {Owen, James E. and Jackson, Alan P.},
	month = oct,
	year = {2012},
	note = {arXiv:1206.2367 [astro-ph]},
	pages = {2931--2947},
}

@article{kubyshkina_grid_2018,
	title = {Grid of upper atmosphere models for 1–40 \textit{{M}}$_{\textrm{⊕}}$ planets: application to {CoRoT}-7 b and {HD} 219134 b,c},
	volume = {619},
	copyright = {https://www.edpsciences.org/en/authors/copyright-and-licensing},
	issn = {0004-6361, 1432-0746},
	shorttitle = {Grid of upper atmosphere models for 1–40 \textit{{M}}$_{\textrm{⊕}}$ planets},
	url = {https://www.aanda.org/10.1051/0004-6361/201833737},
	doi = {10.1051/0004-6361/201833737},
	language = {en},
	urldate = {2025-08-08},
	journal = {Astronomy \& Astrophysics},
	author = {Kubyshkina, D. and Fossati, L. and Erkaev, N. V. and Johnstone, C. P. and Cubillos, P. E. and Kislyakova, K. G. and Lammer, H. and Lendl, M. and Odert, P.},
	month = nov,
	year = {2018},
	pages = {A151},
}

@ARTICLE{Storey1995-uv,
      title     = "Recombination line intensities for hydrogenic {ions-IV}. Total recombination coefficients and machine-readable tables for Z=1 to 8",
      author    = "Storey, P J and Hummer, D G",
      journal   = "Mon. Not. R. Astron. Soc.",
      publisher = "Oxford University Press (OUP)",
      volume    =  272,
      number    =  1,
      pages     = "41--48",
      month     =  jan,
      year      =  1995
}

@BOOK{osterbrock_2006,
        author = {{Osterbrock}, Donald E. and {Ferland}, Gary J.},
        title = "{Astrophysics of gaseous nebulae and active galactic nuclei}",
        year = 2006,
        adsurl = {https://ui.adsabs.harvard.edu/abs/2006agna.book.....O}
}

@article{wordsworth_redox_2018,
	title = {Redox {Evolution} via {Gravitational} {Differentiation} on {Low}-mass {Planets}: {Implications} for {Abiotic} {Oxygen}, {Water} {Loss}, and {Habitability}},
	volume = {155},
	issn = {1538-3881},
	shorttitle = {Redox {Evolution} via {Gravitational} {Differentiation} on {Low}-mass {Planets}},
	url = {https://doi.org/10.3847/1538-3881/aab608},
	doi = {10.3847/1538-3881/aab608},
	language = {en},
	number = {5},
	journal = {The Astronomical Journal},
	author = {Wordsworth, R. D. and Schaefer, L. K. and Fischer, R. A.},
	month = apr,
	year = {2018},
	note = {Publisher: The American Astronomical Society},
	pages = {195},
}

@article{schaefer_predictions_2016,
	title = {{PREDICTIONS} {OF} {THE} {ATMOSPHERIC} {COMPOSITION} {OF} {GJ} 1132b},
	volume = {829},
	issn = {0004-637X, 1538-4357},
	url = {https://iopscience.iop.org/article/10.3847/0004-637X/829/2/63},
	doi = {10.3847/0004-637X/829/2/63},
	language = {en},
	number = {2},
	journal = {The Astrophysical Journal},
	author = {Schaefer, Laura and Wordsworth, Robin D. and Berta-Thompson, Zachory and Sasselov, Dimitar},
	month = oct,
	year = {2016},
	pages = {63},
}

@article{wordsworth_water_2013,
	title = {Water loss from terrestrial planets with {CO2}-rich atmospheres},
	volume = {778},
	issn = {0004-637X, 1538-4357},
	url = {http://arxiv.org/abs/1306.3266},
	doi = {10.1088/0004-637X/778/2/154},
	number = {2},
	journal = {The Astrophysical Journal},
	author = {Wordsworth, Robin and Pierrehumbert, Raymond},
	month = nov,
	year = {2013},
	note = {arXiv:1306.3266 [astro-ph]},
	pages = {154},
}

@article{Verner1996-tw,
  title     = "Atomic data for astrophysics. {II}. New analytic {FITS} for
               photoionization cross sections of atoms and ions",
  author    = "Verner, D A and Ferland, G J and Korista, K T and Yakovlev, D G",
  journal   = "Astrophys. J.",
  publisher = "American Astronomical Society",
  volume    =  465,
  pages     = "487",
  month     =  jul,
  year      =  1996,
  language  = "en"
}

@article{seager_exoplanet_2010,
author = {Seager, Sara},
year = {2010},
month = {08},
pages = {},
title = {Exoplanet Atmospheres: Physical Processes},
journal = {Exoplanet Atmospheres: Physical Processes. By Sara Seager. Princeton University Press, 2010. ISBN: 978-1-4008-3530-0}
}

@article{nakayama_survival_2022,
	title = {Survival of {Terrestrial} {N2}–{O2} {Atmospheres} in {Violent} {XUV} {Environments} through {Efficient} {Atomic} {Line} {Radiative} {Cooling}},
	volume = {937},
	issn = {0004-637X},
	url = {https://doi.org/10.3847/1538-4357/ac86ca},
	doi = {10.3847/1538-4357/ac86ca},
	language = {en},
	number = {2},
	journal = {The Astrophysical Journal},
	author = {Nakayama, Akifumi and Ikoma, Masahiro and Terada, Naoki},
	month = sep,
	year = {2022},
	note = {Publisher: The American Astronomical Society},
	pages = {72},
}

@article{Kempton2023-hm,
  title        = "A reflective, metal-rich atmosphere for {GJ} 1214b from its
                  {JWST} phase curve",
  author       = "Kempton, Eliza M-R and Zhang, Michael and Bean, Jacob L and
                  Steinrueck, Maria E and Piette, Anjali A A and Parmentier,
                  Vivien and Malsky, Isaac and Roman, Michael T and Rauscher,
                  Emily and Gao, Peter and Bell, Taylor J and Xue, Qiao and
                  Taylor, Jake and Savel, Arjun B and Arnold, Kenneth E and
                  Nixon, Matthew C and Stevenson, Kevin B and Mansfield, Megan
                  and Kendrew, Sarah and Zieba, Sebastian and Ducrot, Elsa and
                  Dyrek, Achr{\`e}ne and Lagage, Pierre-Olivier and Stassun,
                  Keivan G and Henry, Gregory W and Barman, Travis and Lupu,
                  Roxana and Malik, Matej and Kataria, Tiffany and Ih, Jegug
                  and Fu, Guangwei and Welbanks, Luis and McGill, Peter",
  journal   = "Nature",
  publisher = "Springer Science and Business Media LLC",
  volume    =  620,
  number    =  7972,
  pages     = "67--71",
  month     =  aug,
  year      =  2023,
  copyright = "https://www.springernature.com/gp/researchers/text-and-data-mining",
  language  = "en"
}

@article{Pass_2023,
    doi = {10.3847/1538-3881/acd6a2},
    url = {https://doi.org/10.3847/1538-3881/acd6a2},
    year = {2023},
    month = {jun},
    publisher = {The American Astronomical Society},
    volume = {166},
    number = {1},
    pages = {16},
    author = {Pass, Emily K. and Winters, Jennifer G. and Charbonneau, David and Irwin, Jonathan M. and Medina, Amber A.},
    title = {Active Stars in the Spectroscopic Survey of Mid-to-late M Dwarfs within 15 pc},
    journal = {The Astronomical Journal},
    }

@article{Peacock2020-bs,
  title        = "{HAZMAT} {VI}: The evolution of extreme ultraviolet radiation
                  emitted from early {M} star",
  author       = "Peacock, Sarah and Barman, Travis and Shkolnik, Evgenya L and
                  Loyd, R O Parke and Schneider, Adam C and Pagano, Isabella
                  and Meadows, Victoria S",
  year         =  2020,
  primaryClass = "astro-ph.SR",
  eprint       = "2005.01687"
}

@article{odert_escape_2018,
	title = {Escape and fractionation of volatiles and noble gases from {Mars}-sized planetary embryos and growing protoplanets},
	volume = {307},
	issn = {00191035},
	doi = {10.1016/j.icarus.2017.10.031},
	journal = {Icarus},
	author = {Odert, P. and Lammer, H. and Erkaev, N. V. and Nikolaou, A. and Lichtenegger, H. I. M. and Johnstone, C. P. and Kislyakova, K. G. and Leitzinger, M. and Tosi, N.},
	month = jun,
	year = {2018},
	note = {arXiv:1706.06988 [astro-ph]},
	pages = {327--346},
}

@misc{cherubim_oxidation_2025,
	title = {An {Oxidation} {Gradient} {Straddling} the {Small} {Planet} {Radius} {Valley}},
	url = {http://arxiv.org/abs/2503.05055},
	doi = {10.48550/arXiv.2503.05055},
	urldate = {2025-03-10},
	publisher = {arXiv},
	author = {Cherubim, Collin and Wordsworth, Robin and Bower, Dan and Sossi, Paolo and Adams, Danica and Hu, Renyu},
	month = mar,
	year = {2025},
	note = {arXiv:2503.05055 [astro-ph]},
}

@article{Moses2013-qx,
  title     = "{COMPOSITIONAL} {DIVERSITY} {IN} {THE} {ATMOSPHERES} {OF} {HOT}
               {NEPTUNES}, {WITH} {APPLICATION} {TO} {GJ} 436b",
  author    = "Moses, J I and Line, M R and Visscher, C and Richardson, M R and
               Nettelmann, N and Fortney, J J and Barman, T S and Stevenson, K
               B and Madhusudhan, N",
  journal   = "Astrophys. J.",
  publisher = "American Astronomical Society",
  volume    =  777,
  number    =  1,
  pages     = "34",
  month     =  nov,
  year      =  2013,
  copyright = "http://iopscience.iop.org/page/copyright",
  language  = "en"
}

@article{Parker1958-wq,
  title     = "Dynamics of the interplanetary gas and magnetic fields",
  author    = "Parker, E N",
  journal   = "Astrophys. J.",
  publisher = "American Astronomical Society",
  volume    =  128,
  pages     = "664",
  month     =  nov,
  year      =  1958,
  language  = "en"
}

@book{Pierrehumbert2010-tf,
  title     = "Principles of planetary climate",
  author    = "Pierrehumbert, Raymond T",
  publisher = "Cambridge University Press",
  month     =  dec,
  year      =  2010,
  address   = "Cambridge, England"
}
\bibliographystyle{aasjournalv7}



\end{document}